# Field-effect transistors made from solution-grown two-dimensional tellurene


Yixiu Wang[1*], Gang Qiu[2, 3*], Ruoxing Wang[1*], Shouyuan Huang[4], Qingxiao Wang[5], Yuanyue Liu[6, 7, 8, 9], Yuchen Du[2, 3], William A. Goddard III[6], Moon J. Kim[4], Xianfan Xu[3, 4], Peide D. Ye[2, 3⋆], Wenzhuo Wu[1, 3⋆]

[1]School of Industrial Engineering, Purdue University, West Lafayette, Indiana 47907, USA

[2]School of Electrical and Computer Engineering, Purdue University, West Lafayette, Indiana 47907, USA

[3]Birck Nanotechnology Center, Purdue University, West Lafayette, Indiana 47907, United States

[4]School of Mechanical Engineering, Purdue University, West Lafayette, Indiana 47907, USA

[5]Department of Materials Science and Engineering, University of Texas at Dallas, Richardson, Texas 75080, USA

[6]The Resnick Sustainability Institute, California Institute of Technology, Pasadena, California 91125, United States

[7]Materials and Process Simulation Center, California Institute of Technology, Pasadena, California 91125, United States

[8]Department of Mechanical Engineering, The University of Texas at Austin, Texas 78712, United States

[9]Texas Materials Institute, and Department of Mechanical Engineering, The University of Texas at Austin, Texas 78712, United States

⋆e-mail: Correspondence and requests for materials should be addressed to W. Z. W (wenzhuowu@purdue.edu) or P. D. Y. (yep@purdue.edu)

*These authors contributed equally to this work.





**Abstract**

The reliable production of two-dimensional crystals is essential for the development of new technologies based on 2D materials. However, current synthesis methods suffer from a variety of drawbacks, including limitations in crystal size and stability. Here, we report the fabrication of large-area, high-quality 2D tellurium (tellurene) using a substrate-free solution process. Our approach can create crystals with a process-tunable thickness, from monolayer to tens of nanometres, and with lateral sizes of up to 100 μm. The chiral-chain van der Waals structure of tellurene gives rise to strong in-plane anisotropic properties and large thickness-dependent shifts in Raman vibrational modes, which is not observed in other 2D layered materials. We also fabricate tellurene field-effect transistors, which exhibit air-stable performance at room temperature for over two months, on/off ratios on the order of $10^6$ and field-effect mobilities of around 700 $cm^2$/Vs. Furthermore, by scaling down the channel length and integrating with high-k dielectrics, transistors with a significant on-state current density of 1 A/mm are demonstrated.


**Main**

The continuing development of two-dimensional materials, be it the exploration of new science[1-3] or the implementation of new technologies[4-8], requires reliable methods of synthesising 2D crystals. Whether current approaches can be scaled up though remains uncertain[9,10] and are restricted by factors such as growth substrates and conditions[11-13], small crystal sizes[14] and the instability of the synthesized materials[11,15,16].



Group VI tellurium (Te) has a unique chiral-chain crystal lattice where individual helical chains of Te atoms are stacked together by van der Waals (vdW) type bonds and spiral around axes parallel to the [0001] direction at the centre and corners of the hexagonal elementary cell[17] (Fig. 1a). Each tellurium atom is covalently bonded with its two nearest neighbours on the same chain. Earlier studies revealed bulk Te has small effective masses and high hole mobilities due to spin-orbit coupling[18]. The lone-pair and anti-bonding orbitals give rise to a slightly indirect bandgap in the infrared regime (~0.35 eV) in bulk Te[19], which has a conduction band minimum (CBM) located at the H-point of the Brillouin zone, and a valence band maximum (VBM) that is slightly shifted from the H-point along the chain direction, giving rise to hole pockets near H-point[20]. When the thickness is reduced, the indirect feature becomes more prominent, as shown by our first-principles calculations (see Methods for computation details). For example, the VBM of 4-layer Te is further shifted to (0.43, 0.34) (in the unit of the surface reciprocal cell), while CBM remains at (1/2, 1/3) (Fig. 1b inset). Accompanied by the shift of VBM, the band gap also increases (Methods, Supplementary Fig. 1), due to the quantum confinement effect, and eventually reaches ~ 1 eV for monolayer Te[21]. Te has other appealing properties, e.g., photoconductivity[22], thermoelectricity[20], and piezoelectricity[23], for applications in sensors, optoelectronics, and energy devices. A wealth of synthetic methods has been developed to derive Te nanostructures[24-26], which favour the 1D form due to the inherent structural anisotropy in Te. Much less is known about the 2D form of Te and related properties.

**Synthesis and Structural Characterization of 2D Tellurene**

In this article, we report a substrate-free solution process for synthesizing large-area, high-quality 2D Te crystals (termed tellurene) with a thickness of a monolayer to tens of nanometres



and a unique chiral-chain vdW structure which is fundamentally different from the layered vdW materials. We use the term X-ene to describe 2D forms of elemental materials without considering the specific bonding[21,27]. The samples are grown through the reduction of sodium tellurite ($Na_2TeO_3$) by hydrazine hydrate ($N_2H_4$) in an alkaline solution at temperatures from 160-200 °C, with the presence of crystal-face-blocking ligand polyvinylpyrrolidone (PVP) (see Methods). Fig. 1c inset shows the optical image of a typical tellurene solution dispersion after reactions at 180 °C for 20 hours when the $Na_2TeO_3$:PVP mole ratio is 52.4:1 (Methods). The 2D Te flakes can be transferred and assembled at large scale into a single layer continuous thin film through a Langmuir-Blodgett (LB) process[28] or a networked continuous thin film[8,29,30] through ink-jet printing (Methods), onto various substrates (Supplementary Fig. 2) for future characterization and device integration. It should be noted that the current LB approaches still lack the sort of desired capabilities in terms of the film continuity and uniformity, compared to other methods such as ink-jet printing[8,29,30], for assembling solution-derived 2D materials into high-performance device. These preliminary results (Supplementary Fig. 2) show potential and warrant more systematic work for future large-scale assembly and applications of solution-derived 2D functional materials.

Individual 2D flakes have edge lengths ranging from 50 to 100 μm, and thicknesses from 10 to 100 nm (Figs. 1d and Supplementary Fig. 3). The structure, composition, and quality of these tellurene crystals have been analyzed by high angle annular dark field scanning transmission electron microscopy (HAADF-STEM), high-resolution transmission electron microscopy (HRTEM), energy dispersive X-ray spectroscopy (EDS), and X-ray diffraction (XRD) (Fig. 1e, f, and Supplementary Fig. 4). Fig. 1e shows a typical atomically-resolved HAADF-STEM image of



tellurene flake (see Methods). The helical chains and a threefold screw symmetry along <0001> are visible (Fig. 1e). The interplanar spacings are 2.2 Å and 6.0 Å, corresponding to Te ($1\bar{2}10$) and (0001) planes[31], respectively. Fig. 1f shows the selective area electron diffraction (SAED) pattern along the [$10\bar{1}0$] zone axis, which is perpendicular to the top surface of the flake. No point defects or dislocations were observed over a large area within single crystals (Supplementary Fig. 4). EDS result confirmed the chemical composition of Te (Supplementary Fig. 4). Similar characterizations and analyses of dozens of 2D Te flakes with different sizes indicate that all samples grow laterally along the <0001> and <$1\bar{2}10$> directions, with the vertical stacking along the <$10\bar{1}0$> directions (Fig. 1g).

**Synthesis Mechanism and Geometric Control for 2D Tellurene**

The controlled PVP concentration is the key for obtaining 2D tellurene. Fig. 2a shows the productivity (see Methods) of tellurene grown at 180 °C with time for a broad range of $Na_2TeO_3$/PVP mole ratios. When a smaller amount of PVP is used, the first 2D structures occur after a shorter reaction time (Fig. 2a, Supplementary Fig. 5). A closer examination of reactions with different PVP concentrations reveals an intriguing morphology evolution in growth products with time. For each PVP concentration, the initial growth products are dominantly 1D nanostructures (Fig. 2b, Supplementary Fig. 5), similar to previous reports[24-26]. After a certain period of reaction, structures possessing both 1D and 2D characteristics start to emerge (Fig. 2b, Supplementary Fig. 5). TEM characterizations indicate that the long axes (showing 1D characteristics) of these flakes are <0001> oriented, and the lateral protruding regions (showing 2D characteristics) grow along the <$1\bar{2}10$> directions, with the {$10\bar{1}0$} facets as the top/bottom surfaces (Supplementary Fig. 6). The 2D regions are enclosed by edges with atomic level step



roughness (Supplementary Fig. 6). These high energy edges are not specific to certain planes during the intermediate states. These structures also have more uneven surfaces compared to 2D tellurene (Supplementary Fig. 6), further manifesting their intermediate nature. Finally, the ratio of 2D tellurene flakes which have a straight $\{1\bar{2}10\}$ edge increases with a reduction in 1D and intermediate structures (Supplementary Fig. 5) and reaches a plateau after an extended growth, e.g. ~30 hours (Fig. 2a, Supplementary Fig. 5). The growth with a lower level of PVP has a smaller final productivity (Fig. 2a, Supplementary Fig. 5). The observed morphology evolution suggests that the balance between the kinetic and thermodynamic growth dictates the transformation from 1D structures to 2D forms (Fig. 2b). In the initial growth, PVP is preferentially adsorbed on the $\{10\bar{1}0\}$ surfaces of the nucleated seeds[26], which promotes the kinetic-driven 1D growth (Supplementary Fig. 5). When the reaction continues, $\{10\bar{1}0\}$ surfaces of the formed structures would become partially covered due to the insufficient PVP capping. Since $\{10\bar{1}0\}$ surfaces have the lowest free energy in tellurium[32], the growth of $\{10\bar{1}0\}$ surfaces along the $<1\bar{2}10>$ direction significantly increases through the thermodynamic-driven assembly, giving rise to the observed intermediate structures. The enhanced growth along the $<1\bar{2}10>$ directions together with the continued $<0001>$ growth (Supplementary Fig. 6) leads to the formation of 2D tellurene (Supplementary Fig. 5, Fig. 2b).

The sizes and thicknesses of tellurene can also be effectively modulated by controlling the ratio between sodium tellurite and PVP (Fig. 2c, Supplementary Fig. 7). The width of tellurene monotonically decreases with the reduction of PVP level; the thickness is minimized when a medium level of PVP is used (e.g., $Na_2TeO_3$/PVP ratio ~52.4/1, Group #12 in Fig. 2c and Supplementary Fig. 7), and increases with both the increase and decrease of PVP (Fig. 2c). With



a small amount of PVP, the solution is supersaturated with Te source, and homogeneous nucleation of Te can occur in large scale, consuming resource for subsequent growth. As a result, the Ostwald ripening of Te nuclei is shortened, and the final tellurene crystals have smaller sizes compared to samples grown at higher PVP concentrations. The low PVP level also leads to more significant growth along thickness directions. On the other hand, when the PVP level is high, the fewer nucleation events allow the sufficient supply of Te source for subsequent growth, leading to the increased width and thickness. Also, the productivity of tellurene increases with the reaction temperature from 160 °C to 180 °C (Supplementary Fig. 8). This is likely because higher temperature promotes the forward reaction rate in the half reaction of endothermic hydrazine oxidation (see Supplementary Notes). However, when temperature increases from 180 ℃ to 200 ℃, the possible breaking of the van der Waals bonds between Te chains by the excessive energy could lead to the saturated productivity.

Tellurene crystals with a thickness smaller than 10 nm to ultimately monolayer structure can be further derived through a solvent-assisted post-growth thinning process (see Methods). The thickness of tellurene decreases with time after acetone is introduced into the growth solution (Supplementary Fig. 9). After 6 hours, the average thickness of tellurene is reduced to ~ 10 nm, with the thinnest flake down to 4 nm thick (~ 10 layers) (Supplementary Fig. 9). Due to the poor solubility in acetone, PVP molecules tend to desorb from the tellurene and undergo aggregation[33], giving rise to the sediment of tellurene over the time in acetone (Supplementary Fig. 9). Lacking the protection of PVP, the tellurene surfaces get exposed and react with the alkaline growth solution (pH ~ 11.5)[34], leading to the reduced thickness. We have also performed control experiments using other types of solvents in the growth solution



(Supplementary Fig. 10), the results of which suggest that PVP solubility in the solvent significantly affects the above process. Large-area (up to 100-μm in lateral dimensions) tellurene crystals with monolayer, bi-layer, tri-layer and few-layer thicknesses can be further obtained (Fig. 2d, Supplementary Fig. 11), by controlling the pH values of the tellurene dispersion solution in the above post-growth thinning process (see Methods, Fig. 2d, Supplementary Fig. 12).

**Thickness- and Angle-dependent Raman Spectra**

These high-quality ultrathin tellurene crystals with controlled thicknesses provide an ideal system to explore their intrinsic properties in the 2D limit. We first characterized the optical properties of as-synthesized tellurene with a wide range of thickness (from a monolayer to 37.4 nm) by angle-resolved polarized Raman spectroscopy at room temperature (see Methods). The incident light comes in along the [10$\bar{1}$0] direction and is polarized into the [0001] helical chain direction of the tellurene. The Raman spectra of tellurene samples with different thickness (Fig. 3a) exhibits three main Raman-active modes, with one A-mode and two E-modes which correspond to the chain expansion in basal plane, bond-bending around [1$\bar{2}$10] direction and asymmetric stretching mainly along [0001] helical chain[35], respectively (Supplementary Fig. 13). For the 2D Te samples thicker than 20.5 nm, three Raman-active modes locating at 92 cm$^{-1}$ ($E_1$ transverse (TO) phonon mode), 121 cm$^{-1}$ ($A_1$-mode) and 143 cm$^{-1}$ ($E_2$-mode) were identified (Fig. 3a), which agrees well with previous observations in bulk and nanostructured tellurium[36-38], indicating that although these thicker crystals possess 2D morphology, their symmetric properties can still be characterized as bulk. The appreciable effective dynamic charge induced for the $E_1$ mode in tellurium leads to a split of $E_1$ doublets at



92 cm$^{-1}$ and 105 cm$^{-1}$ for transverse (TO) or longitudinal (LO) phonons, respectively[37]. The absence of E$_1$ (LO) mode in our observed results for 2D Te thicker than 20.5 nm, similar to previous reports on bulk and nanostructured tellurium[36-38], may be attributed to the different signs in the deformation potential and electro-optic contribution to the Raman scattering tensor, which gives rise to the cancellation if both contributions have the same magnitude[39]. As the thickness decreases from 20.5 nm to 9.1 nm (Fig. 3a), the deformation potential in tellurene lattice increases while the electro-optic effect weakens[40], leading to the appearance of E$_1$ (LO) mode in the Raman spectra for 2D Te crystals with intermediate thickness. When the 2D Te's thickness further reduces (smaller than 9.1 nm in Fig. 3a), the degeneracy in the E$_1$ TO and LO modes occurs with peak broadening, possibly due to the intra-chain atomic displacement, the electronic band structure changes and the symmetry assignments for each band[41], all of which are affected by the sample thickness.

When tellurene's thickness decreases, there are significant blue-shifts for both A$_1$ (shift to 136 cm$^{-1}$ for monolayer) and E$_2$ modes (shift to 149 cm$^{-1}$ for monolayer) (Fig. 3a). The hardened in-plane E$_2$ vibration mode in thinner tellurene, similar to reported observations for black phosphorus[42] and MoS$_2$[43,], may be attributed to the enhanced interlayer long-range Coulombic interactions when thinned down. The observed blue-shift for the A$_1$ mode in 2D Te, in strong contrast to 2D layered vdW materials which usually witness red-shift for the out-of-plane vibration mode when thinned down[15,42,43], is thought to be closely related to the unique chiral-chain vdW structure of tellurene. When thinned down, the lattice deformation within the 2D plane gave rise to the attenuated inter-chain vdW interactions and enhanced intra-chain covalent interactions in the individual tellurene layer, leading to more effective restoring forces



on tellurium atoms and hence hardened out-of-plane $A_1$ vibration mode (Supplementary Fig. 13). Such unique structure of tellurene also results in the giant thickness-dependent shift in Raman vibrational modes, which is unseen in 2D layered vdW materials[42-44]. The interaction between the substrate ($SiO_2$/Si) and 2D Te flakes could also contribute to the hardened $A_1$ and $E_2$ modes[36]. The stiffening of vibrational modes in monolayer tellurene (Fig. 3a) is consistent with its structure reconstruction where extra bonds are formed between neighbouring chains in the single layer tellurium[21,27,45].

Reduced in-plane symmetry in the chiral-chain vdW structure of tellurene indicates a strong in-plane anisotropy for its material properties. We further characterized the anisotropic optical properties of as-synthesized tellurene with three different thicknesses (28.5 nm, 13.5 nm, and 9.7 nm) by angle-resolved polarized Raman spectroscopy at room temperature (see Methods). By rotating the tellurene flakes in steps of 15°, we observed the changes in the angle-resolved Raman peak intensities (Fig. 3b, Supplementary Figs. 14a, 15a). We extracted the peak intensities of different modes by fitting with Lorentz function and plotted them into the corresponding polar figures (Figs. 3c-f, Supplementary Figs. 14b-e, 15b-d) (see Methods). While all the modes change in intensity with the polarization angle, we find that the peak for the $A_1$ mode in all samples exhibits the largest sensitivity to the relative orientation between the [0001] direction and the polarization of the excitation laser (middle panels of Fig. 3b, Supplementary Figs. 14a, 15a). It is worth to note that the direction of maximum intensity in $A_1$ polar plot changes with the sample thickness. More specifically, $A_1$ polar plots for the 13.5 nm and 28.5 nm samples show the maximum intensity at 90° and 270° along the [1$\bar{2}$10] direction (Fig. 3e, Supplementary Fig. 15c). However, when the thickness decreases to 9.7 nm,



the maximum intensity direction switches to 0° and 180° along the [0001] direction

(Supplementary Figs. 14d). A similar phenomenon also occurs for $E_1$-LO modes in the 13.5 nm

and 9.7 nm samples (Fig. 3d, Supplementary Figs. 14c). Such thickness-dependent anisotropic

Raman scattering could be attributed to the different absorption spectral range in [0001]

and [1$\bar{2}$10] directions[46] and anisotropic interference effect[41]. The angle-resolved Raman

results also confirm that the helical Te atom chains in the as-synthesized tellurene are oriented

along the growth direction of the tellurene flake, which matches the TEM results (Fig. 1e).

**Device Implementation of Tellurene Field-effect Transistors**

Finally, we explored the electrical performance of tellurene field-effect transistors (FETs) to

demonstrate its great potential for logic electronics application. Back-gate devices were

fabricated on high-k dielectric substrates and source/drain regions were patterned by

electron beam lithography with the channel parallel to the [0001] direction of tellurene (details

see Methods). We chose Pd/Au (50/50 nm) as metal contacts since Pd has relatively high work

function that can reduce the contact resistance in *p*-type transistors[47-49]. Long channel devices

were first examined (channel length 3 μm) where the contact resistance is negligible, and the

transistor behavior is dominated by intrinsic electrical properties of channel material. Fig. 4a

shows the transfer curve of a typical 7.5-nm-thick 2D Te FET measured at room temperature.

The device exhibits *p*-type characteristics with slight ambipolar transport behavior due to its

narrow bandgap nature, with large drain current over 300 mA/mm (see Supplementary Fig. 23)

and high on/off ratio on the order of ~$10^5$. The *p*-type behavior originates from the high level of

Te valance band edge, as shown by our first-principles calculations (Supplementary Fig. 16).

Meanwhile, the process-tunable thickness of tellurene allows the modulation of electrical



performance in tellurene transistors. Overall, important metrics of tellurene-based transistors such as on/off ratio, mobility, and on-state current level are superior or comparable to transistors based on other 2D materials[11,15,16,50]. We further investigated the thickness dependence of two key metrics of device performance, namely on/off ratios and field-effect mobilities, for more than 50 2D Te long channel devices with flake thicknesses ranging from over 35 nm down to a monolayer (~0.5 nm), to elucidate the transport mechanism of 2D Te FETs (Fig. 4b). The linear behavior of the output curves in the low $V_{ds}$ region (Supplementary Fig. 23) suggests that the contact resistance is low (see Supplementary Notes and Supplementary Fig. 19 for the extracted contact resistance) which ensures the soundness of the field-effect mobility extraction from the slope of the linear region of the transfer curves (see Methods). The field-effect mobilities of 2D Te transistors peak with ~700 cm$^2$/Vs at room temperature at around 16 nm thickness and decreases gradually with the further increase of the thickness. The transfer curves of devices with thin thickness of 2.8 nm (~ 6 layers), 1.7 nm (~ 4 layers), 1.0 nm (bi-layer) and 0.5 nm (monolayer) are included in Supplementary Fig. 24. A benchmark comparison with black phosphorus, which is also a narrow bandgap p-type 2D material, shows that solution-synthesized 2D Te has ~ 2-3 times higher mobility than black phosphorus when the same device structure, geometry, and mobility extraction method are adopted[15] (Supplementary Fig. 21). This thickness-dependence is similar to other layered materials that experience screening and interlayer coupling[15,16] (Supplementary Notes and Supplementary Fig. 20). The field-effect mobility is also affected by the contribution of the carriers from layers near the semiconductor-oxide interface. Thinner samples are more susceptible to the charge impurities at the interface and surface scattering, which explains the



decrease in the mobility of the few-layer tellurene transistors. We expect to be able to improve the mobility of tellurene through approaches such as improving interface quality with high-k dielectric[51] or h-BN encapsulation to reduce the substrate phonon scattering and charge impurity. For bi-layered tellurene transistors, external field-effect mobility is dropped to ~1 cm$^2$/Vs. This is due to bandgap increasing in few-layer tellurene to form much higher Schottky barrier, which may introduce large deviation in extracting mobility due to the drastically increased contact resistance. Because of the reduced gate electrostatic control in thicker flakes, the thickness-dependent on/off ratios (see Supplementary Notes) steeply decrease from ~10$^6$ to less than 10 once the crystal thickness approaches the maximum depletion width of the films, with a trend similar to other reported narrow bandgap depletion-mode 2D FETs[15,16].

The in-plane anisotropic electrical transport properties were also studied at room temperature. To minimize flake-to-flake variation and geometric non-ideality, we applied dry-etching method (see Methods) to trim two identical rectangles from the same 2D Te flake. One of the rectangles was aligned along the 1D atomic chain [0001] direction and the other along [1$\bar{2}$10] direction (Fig. 4c inset). Long channel FETs (channel length 8 μm) were fabricated to minimize contact influence and manifest the intrinsic material properties. The extracted field-effect mobilities along these two primary directions from seven 2D Te samples exhibit an average anisotropic mobility ratio of 1.43±0.10 (Fig. 4c). A typical set of results measured from a 22-nm-thick sample is shown in Supplementary Fig. 22. This anisotropic ratio in mobility is slightly lower than that reported for bulk tellurium[52], possibly due to the enhanced surface scattering in our ultrathin Te samples. Our first-principle calculations show a similar degree of



anisotropy in the effective masses along these two orthogonal directions (Fig. 1b, 0.32 $m_0$ perpendicular to the chain and 0.30 $m_0$ along the chain, see Methods).

Great air-stability was also demonstrated in tellurene transistors with different flake thickness. Electrical performance of a 15-nm-thick transistor was monitored after being exposed in air for two months without any encapsulation, as shown in Fig. 4d. No significant degradation was observed in the same device during two-month period, except slight threshold voltage shift probably coming from sequential measurement variation. We further demonstrated that such good air-stability is valid for almost the entire thickness range from thick flakes down to 3 nm (See Supplementary Fig. 25). For even thinner flakes, the thin films are no longer conducting after first couple of days.

More strikingly, by scaling down the channel length and integrating with our ALD-grown high-k dielectric, we achieved record high drive current of over 1 A/mm at relatively low $V_{ds}$=1.4 V. Fig. 5a and 5b represent I-V curves of a short channel device with channel length of 300 nm fabricated on an 11-nm-thick Te flake. The on/off ratio at small drain bias ($V_{ds}$=-0.05 V) is over $10^3$, which is still a decent value, considering its narrow bandgap of ~0.4 eV (see Supplementary Fig. 1). The off-state performance is slightly deteriorated at large drain voltage ($V_{ds}$=-1 V, pink circles in Fig. 5b) due to the short channel effect. Large drain voltage reduces the barrier height for electron branch and electron current is unhindered, which is also reflected in the upswing of drain current at large $V_{ds}$ in output curve (Fig. 5a). Such effect is common in narrow-bandgap short channel devices[53,54] and can be mediated through proper contact engineering[54]. Fig. 5c shows the relationship between two transistor key parameters, on/off ratio and maximum drain current, of over 30 devices with different channel thickness. Generally speaking, a short



channel device with flake thickness around 7-8 nm offers the best performance with on/off ratio ~$10^4$ and maximum drain current > 600 mA/mm. It is also worth mentioning that the maximum drain current we achieved is 1.06 A/mm, with several devices exceeding 1 A/mm, which is so far the highest value among all the two dimensional material transistors to our best knowledge[53,55,56]. This number is also comparable to that of conventional semiconductor devices.

**Conclusions**

We have developed a simple, low-cost, solution-based approach for the scalable synthesis and assembly of 2D Te crystals. These high-quality 2D Te crystals have high carrier mobility and are air-stable (measured up to two months). Our prototypical 2D Te device shows a good all-around figure of merits (Supplementary Fig. 26, Supplementary Table 1) compared to existing 2D materials and record-high on-state current capacity. Our approach has the potential to produce stable, high-quality, ultrathin semiconductors with a good control of composition, structure, and dimensions, opening up opportunities for applications in electronics, optoelectronics, energy conversion, and energy storage. 2D Tellurene, as a chiral-chain van der Waals solid, adds a new class of nanomaterials to the large family of 2D crystals.



**Methods**

**Synthesis of 2D tellurene crystals**

In a typical procedure, analytical grade $Na_2TeO_3$ (0.00045 mol) and a certain amount of poly(-vinyl pyrrolidone) was put into double distilled water (33 ml) at room temperature under magnetic stirring to form a homogeneous solution. The resulting solution was poured into a Teflon-lined stainless-steel autoclave, which was then filled with an aqueous ammonia solution (25%, w/w%) and hydrazine hydrate (80 %, w/w%). The autoclave was sealed and maintained at the reaction temperature for a designed time. Then the autoclave was cooled to room temperature naturally. The resulting silver-gray, solid products were precipitated by centrifuge at 5000 rpm for 5 minutes and washed 3 times with distilled water (to remove any ions remaining in the final product).

**Langmuir-Blodgett (LB) transfer of tellurene**

The hydrophilic 2D Te nanoflake monolayers can be transferred to various substrates by the Langmuir-Blodgett (LB) technique. The washed nanoflakes were suspended in a mixture solvent made of N, N-dimethylformamide (DMF) and $CHCl_3$ (e.g., in the ratio of 1.3:1). Then, the mixture solvent was dropped into the deionized water. Too much DMF will result in the falling of 2D Te in the water. It is difficult to mix the DMF, $CHCl_3$ and 2D Te when $CHCl_3$ is too much. After the evaporation of the solvent, a monolayer assembly of 2D Te flakes was observed at the air/water interface. Then we can transfer the monolayer assembly of 2D Te onto the substrates.

**First-Principles Calculations**

Density Functional Theory calculations were performed using the Vienna Ab-initio Simulation Package (VASP)[57] with projector augmented wave (PAW) pseudopotentials[58]. We used 500 eV for



the plane-wave cutoff, 5x5x1 Monkhorst-Pack sampling, and fully relaxed the systems until the final force on each atom was less than 0.01 eV/Å. The PBE exchange-correlation functional is used for relaxation of the system, and the HSE functional is employed to calculate the band gaps (Fig. S1) and the band edge levels (Fig. S10). We find a significant structural reconstruction for monolayer Te, in agreement with reported result[21]. While for bilayer and thicker Te, the structure is similar to that of bulk Te. Our calculations show a lattice parameter of 4.5 Å and 6.0 Å for multilayers, in agreement with experiments. The adsorption of O on bilayer Te and P is modeled by using 4x3 cell (see Fig. S12).

**Structural characterization**

The morphology of the ultrathin tellurene crystals was identified by optical microscopy (Olympus BX-60). The thickness was determined by AFM (Keysight 5500). High-resolution STEM/TEM imaging and SAED has been performed using a probe-corrected JEM-ARM 200F (JEOL USA, Inc.) operated at 200 kV and EDS has been collected by an X-MaxN100TLE detector (Oxford Instruments). In HAADF-STEM mode, the convergence semi-angle of electron probe is 24 mrad, and the collection angle for ADF detector was set to 90-370 mrad.

**Determination of tellurene productivity**

To quantify the ratio of 2D tellurene flake in the products, we measured all the products in the same process as follows: the freshly prepared 2D tellurene solution (1 mL) was centrifuged at 5000 rpm for 5 minutes after adding acetone (2 mL) and washed with alcohol and double distilled water twice. Then the 2D tellurene flakes were dispersed into 3 mL double distilled water. After that, we dropped 100 µL dispersed solution onto the 1×1 cm$^2$ SiO$_2$/Si substrate. After the water evaporation, we used the optical microscope to record several images randomly covering



the 5×5 mm$^2$ area. In the end, we analyze the areas covered by 2D tellurene by ImageJ, public domain, Java-based image processing program developed at the National Institutes of Health. In our case, we define the productivity as the ratio of the 2D tellurene area in the entire image.

**Solvent-assisted post-growth thinning process**

For thinning process using the alkaline growth solution, the as-synthesized 2D tellurene solution (1 mL) was mixed with acetone (3 mL) at the room temperature. After a specific time (e.g., 6 hours), the thin 2D tellurene can be obtained by centrifuge at 5000 rpm for 5 minutes. After doing the LB process, the 2D tellurene can be transferred onto the substrate.

For thinning process using tellurene solution with controlled pH values, the suspension of as-synthesized 2D Te (1 mL) was centrifuged one time with the addition of 3 mL DI water. Then, the 2D Te was dispersed into a solution of 1 mL of NaOH and 3 mL of acetone. The concentration of NaOH was varied to control the pH value of the above 4 mL solution. After that, the above solution was kept at room temperature for 2-10 hours. Finally, the thinned tellurene samples were precipitated by centrifuge.

**FET Device fabrication and characterization**

High-k dielectric stack consisting of 20 nm hafnium zirconium oxide (Hf$_{0.5}$Zr$_{0.5}$O$_2$) and 2 nm Al$_2$O$_3$ was first deposited by atomic layer deposition (ALD) onto heavily doped n++ silicon wafer. Upon transferring the tellurene flakes onto the substrate, source and drain regions were patterned by electron beam lithography (EBL). We chose 50/50 nm Pd/Au for contact metal since Pd has relatively high work function which benefits the p-type transistors by reducing Schottky contact resistance. The electrical measurements were performed using Keithley 4200A semiconductor characterization system.



By plugging numbers into the formula: $\mu_{FE} = g_m L / W C_{ox} V_{ds}$, where $g_m, L, W$ and $C_{ox}$ are transconductance, channel length, channel width and gate oxide capacitance, we can derive the field-effect mobilities for the tellurene transistors. Field-effect mobilities extracted from devices fabricated on tellurene crystals with various thicknesses are displayed in Fig. 4b. Devices for anisotropic transport measurement were first patterned into two perpendicular rectangles along the two principle in-plane directions of tellurene with EBL and dry-etched into desired shapes with BCl$_3$ and argon plasma. The rest of fabrication process follows the same route as before.

**Raman Spectra**

Angle-resolved Raman Spectra were measured at room temperature. The crystal symmetry of Te renders one A$_1$ mode, one A$_2$ mode (Raman-inactive), and two doublet E modes at Γ point of Brillouin zone. Raman signal was excited by 633nm He-Ne laser. The incident light comes in along [-1010] direction which is perpendicular to the Te flake surface and was polarized into [0001] direction, which is parallel to spiral atom chains and we denote this configuration as 0°. A linear polarizer was placed in front of the spectrometer to polarize reflected light into the same direction with incident light. Polarized Raman can eliminate all the other superimposed Raman signals and manifests a clear trace of angle-dependent Raman spectrum evolution. Then by rotating the Te flake, we observed angle-resolved Raman peak intensity change, as shown in Fig. 3a. We extracted the peak intensities of different modes by fitting with Lorentz function and plotted them into polar figures (Fig. 3b-f). These angle dependent behaviors were then fitted by calculating matrix multiplication: $\boldsymbol{e_i} \times \boldsymbol{R} \times \boldsymbol{e_r}$, where $\boldsymbol{e_i}$ and $\boldsymbol{e_r}$ are unit vectors of incident the and reflected light direction and $\boldsymbol{R}$ is the Raman tensor of corresponding Raman modes[2]. The



angle-resolved Raman results confirm that the helical Te atom chain is indeed along the long axis of the Te flake, which matches our previous TEM results.

**Data Availability Statement**

The data that support the plots within this paper and other findings of this study are available from the corresponding author upon reasonable request.

**Online Content** Supplementary notes and Extended Data display items are available in the online version of the paper; references unique to these sections appear only in the online paper.

**Acknowledgements** W. Z. W. acknowledge the College of Engineering and School of Industrial Engineering at Purdue University for the startup support. W. Z. W. was partially supported by a grant from the Oak Ridge Associated Universities (ORAU) Junior Faculty Enhancement Award Program. Part of the solution synthesis work was supported by the National Science Foundation under grant no. CMMI-1663214. P. D. Y. was supported by NSF/AFOSR 2DARE Program, ARO, and SRC. Q. W. and M. J. K. were supported by the Center for Low Energy Systems Technology (LEAST) and Center for South West Academy of Nanoelectronics (SWAN). Y. L. thanks support from Resnick Prize Postdoctoral Fellowship at Caltech, and the startup support from UT Austin. Y. L. and W. A. G. were supported as part of the Computational Materials Sciences Program funded by the U.S. Department of Energy, Office of Science, Basic Energy Sciences, under Award Number DE-SC00014607. This work used computational resources of NREL (sponsored by DOE EERE), XSEDE (NSF ACI-1053575), NERSC (DOE DE-AC02-05CH11231), and the Texas Advanced Computing Center (TACC) at UT Austin. We would like to thank Dr. Fengru Fan for the helpful discussions.


**Author Contribution Statement** W. Z. W. and P. D. Y. conceived and supervised the project. W. Z. W., P. D. Y., Y. X. W., and G. Q. designed the experiments. Y. X. W. and R. X. W. synthesized the material. G. Q. and Y. X. W. fabricated the devices. G. Q. and Y. C. D. performed the electrical and optical characterization. S. Y. H. and Y. X. W. performed the Raman measurement under the supervision of X. F. X. Q. W. and M. J. K. performed the TEM characterization. Y. L.



carried out the first-principles calculations under the supervision of W. A. G. Y. X. W. and G. Q. conducted the experiments. W. Z. W., P. D. Y., Y. X. W., G. Q., and R. X. W. analyzed the data. W. Z. W. and P. D. Y. wrote the manuscript. Y. X. W., G. Q., and R. X. W. contributed equally to this work. All authors have discussed the results and commented on the paper.

**Competing financial interests** The authors declare no competing financial interests.

**Author Information** Supplementary information is available in the online version of the paper. Reprints and permissions information is available online at www.nature.com/reprints. Readers are welcome to comment on the online version of the paper. Correspondence and requests for materials should be addressed to W. Z. W. (wenzhuowu@purdue.edu) or P. D. Y. (yep@purdue.edu) .



**Figures**

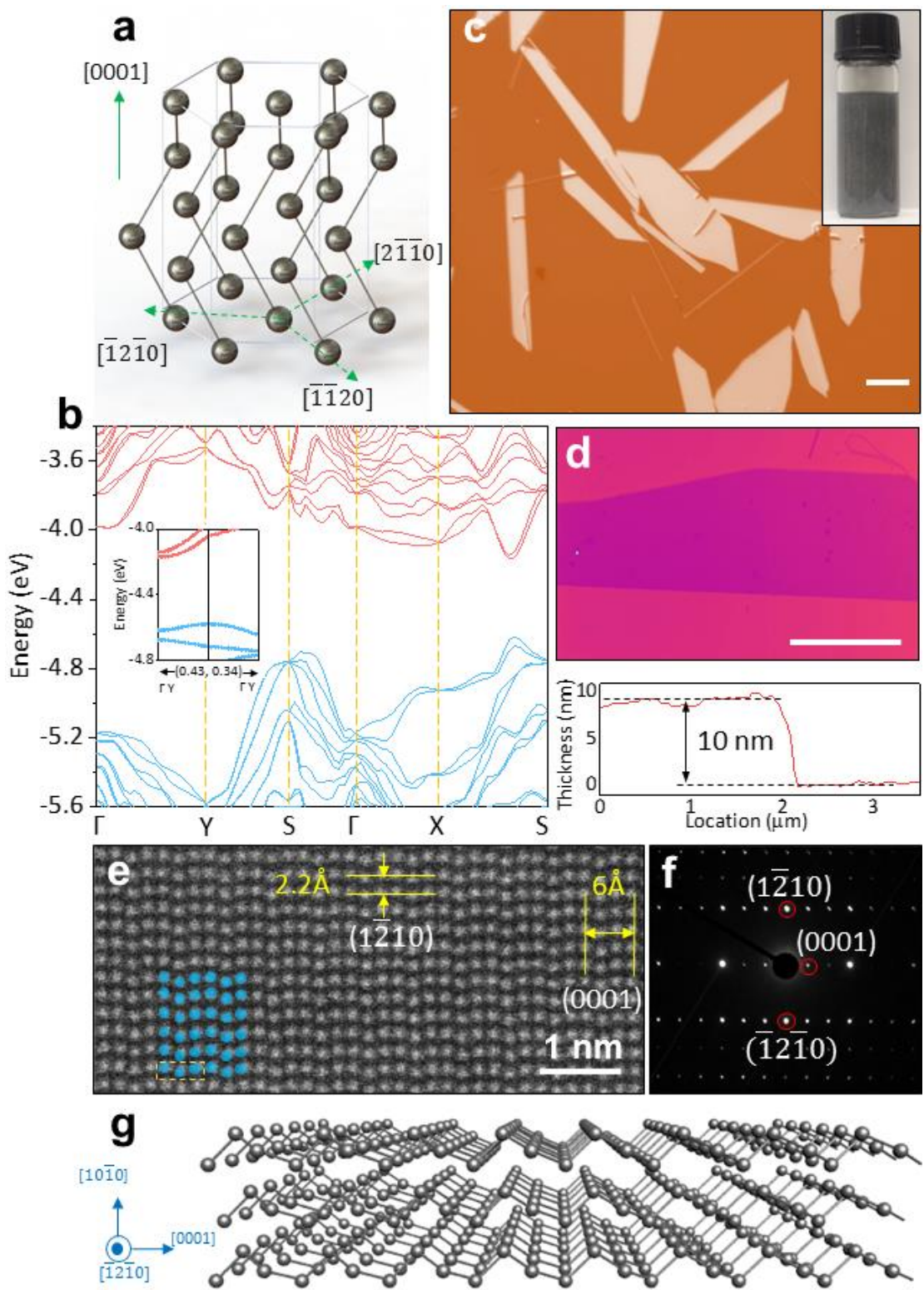

**Figure 1 | Solution-grown large-area 2D Te and material characterization. a**, Atomic structure of tellurium. **b,** The band structure of 4-layer Te, calculated by using PBE functional. Valence bands



are shown in blue, and conduction bands are in red. The inset shows the local band structure near the VBM. Γ: (0,0); X: (0.5, 0); Y: (0, 0.5); S: (0.5, 0.5); all in the units of the surface reciprocal lattice vectors. See Supplementary Information for further information of the electronic structure. **c,** Optical image of solution-grown tellurene flakes. Inset: Optical image of the tellurene solution dispersion. The scale bar is 20 μm. **d,** AFM image of a 10-nm 2D Te flake. The scale bar is 30 μm. **e,** (HAADF)-STEM image of tellurene. The false-colored (in blue) atoms are superimposed to the original STEM image for highlighting the helical structure. **f,** Diffraction pattern of tellurene. **g,** 3D illustration of tellurene's structure.



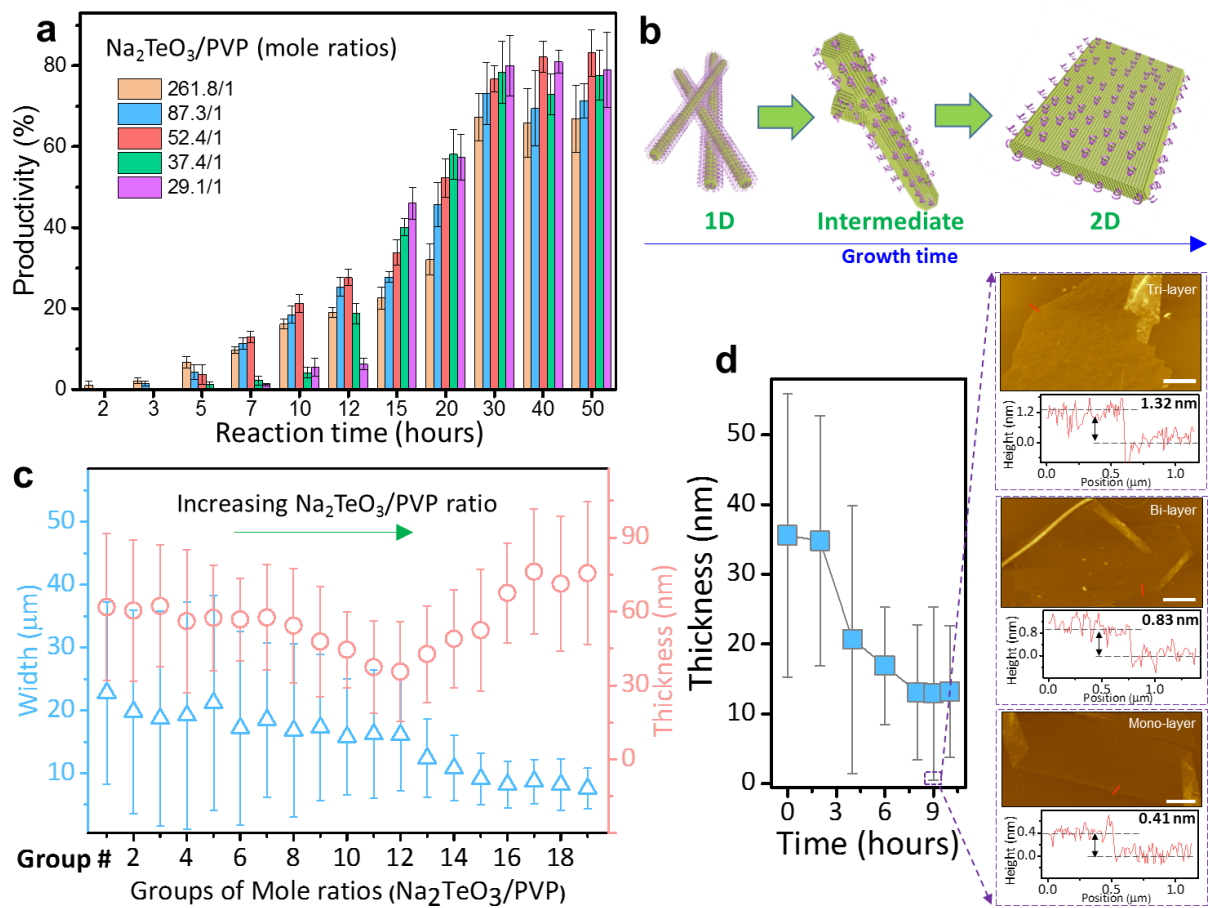

**Figure 2 | Solution processing for tellurene. a**, Growth outcome for different PVP concentrations with different reaction time. Mean values from 5 technical replicates are indicated. Error bars represent s.d. **b,** Morphology evolution from 1D Te structures to 2D Te. **c,** Thickness and width modulation of 2D Te. Mean values from 8 technical replicates are indicated. Error bars represent s.d. **d,** Post-growth thinning process with solution pH = 10.5 for obtaining ultrathin few-layer and monolayer tellurene. The scale bar is 5 μm. Mean values from 8 technical replicates are indicated. Error bars represent s.d.



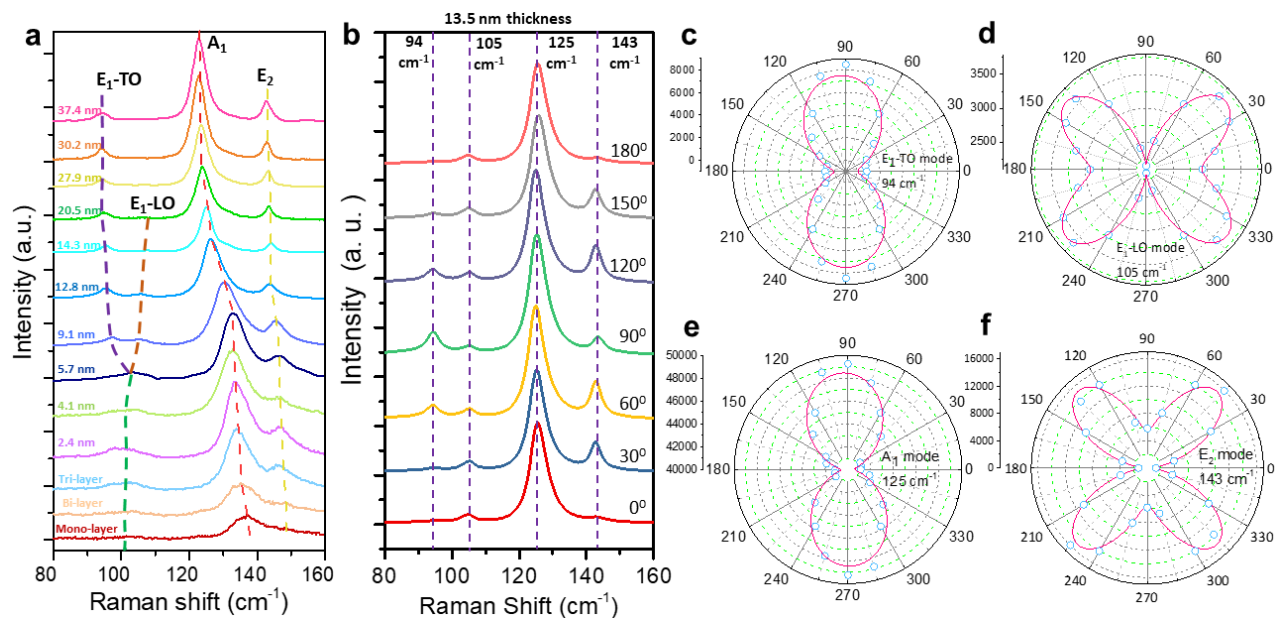

**Figure 3 | Angle-resolved Raman Spectra for 2D tellurene with different thicknesses. a,** Raman spectra for 2D Te with different thicknesses. **b-f,** Angle-resolved Raman spectra for a 13.5-nm-thick flake. **b,** Evolution with angles between crystal orientation and incident laser polarization. **c-f,** Polar figures of Raman Intensity corresponding to $A_1$ and two E modes located at 94 ($E_1$-TO), 105 ($E_1$-LO), 125 ($A_1$), and 143 ($E_2$) cm$^{-1}$. The fitting curves are described in Supplementary methods.



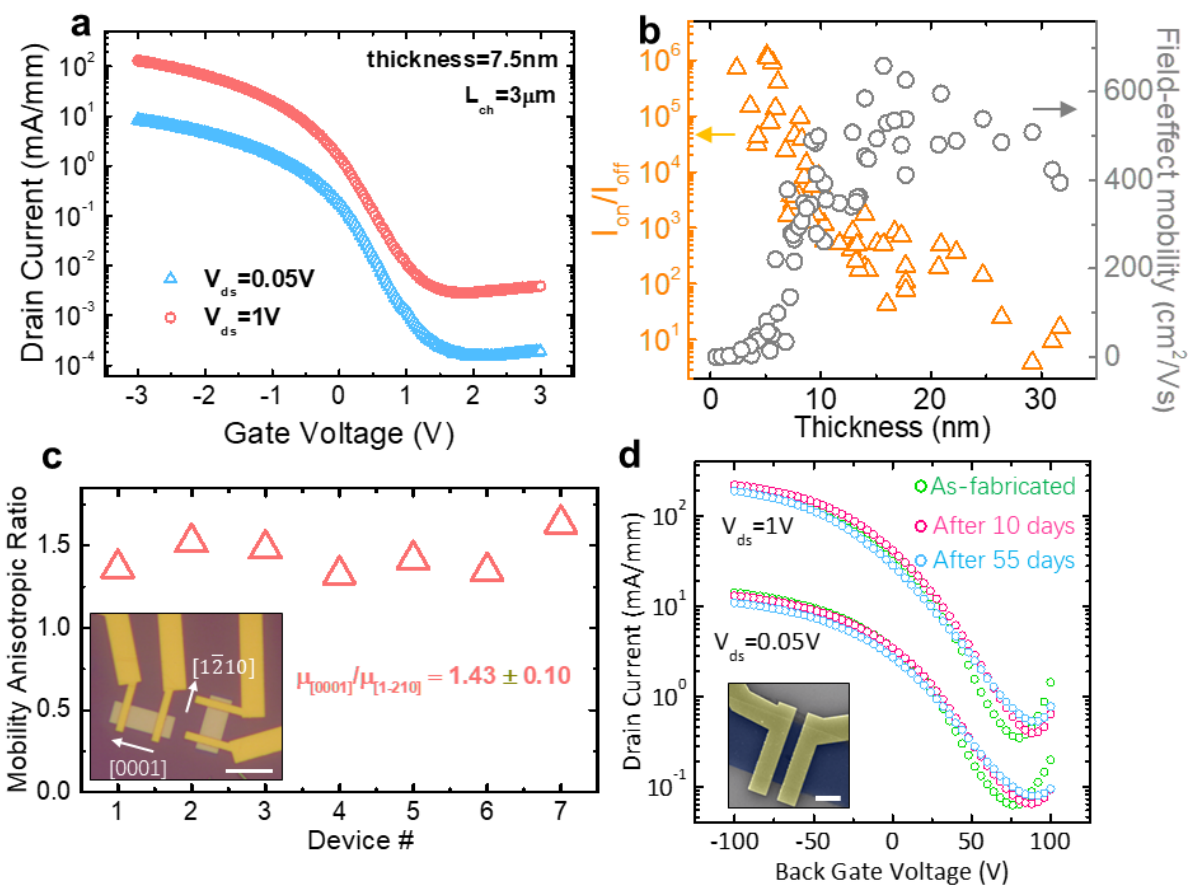

**Figure 4 | 2D tellurene FET device performance. a,** Transfer curve of a typical long channel 2D tellurene transistor with a thickness of 7.5 nm. **b,** Thickness-dependent on/off ratio (orange triangles) and field-effect mobility (gray circles) for 2D Te transistors. The non-monotonic dependence of mobility on thickness can be fitted using Thomas-Fermi screening model (see Supplementary Notes and Supplementary Figure S20). **c,** Anisotropic mobility measurements along [0001] and [1$\bar{2}$10] directions. (Inset: Optical image of a typical device for anisotropic transport measurement. The scale bar is 10 μm.) **d,** Transfer curves of a typical 2D tellurene transistor with a thickness of 15 nm and stability measured for 55 days. (Inset: False-colored SEM image of a tellurene transistor. The scale bar is 10 μm.)



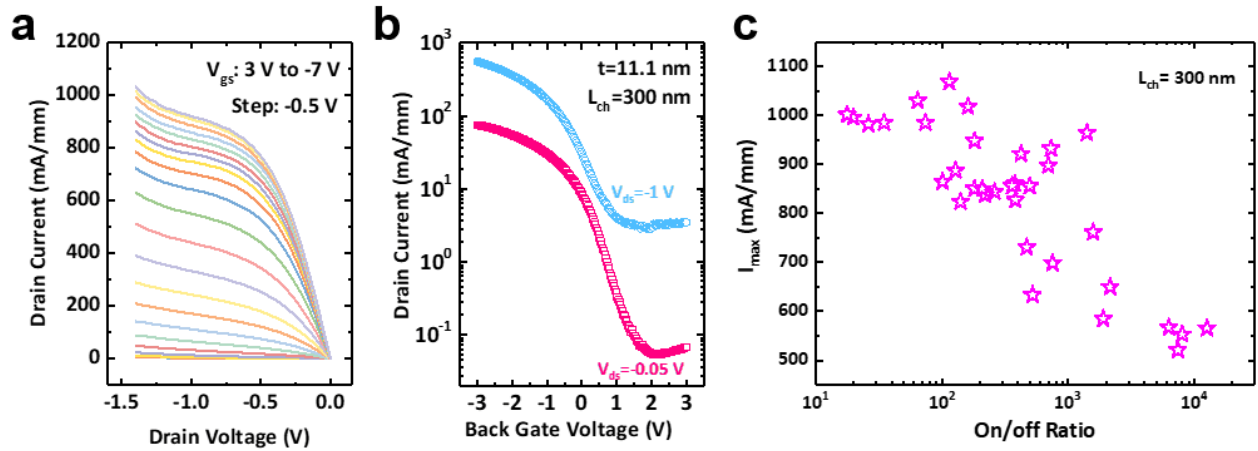

**Figure 5 | High on-state current density in short-channel tellurene devices. a,** The output and **b,** transfer curves of an 11.1-nm-thick tellurene transistor with 300-nm channel. **c,** Trade-off between on/off ratio and maximum drain current measured in over 30 short-channel devices with the same geometry and dimension as in **a** and **b**. The maximum drain current of several devices surpassed 1 A/mm, which is so far the highest value among all 2D material devices.



1. **Supplementary Notes**

   **Supplementary Note 1: Discussion on the temperature effect**

   The tellurene synthesis is also influenced by the reaction temperature. The reaction was carried out through the reduction of Na$_2$TeO$_3$ by hydrazine in alkaline solution. The reaction equation can be written as follow[9]:

   $$TeO_3^{2-} + 3H_2O + 4e^- \rightarrow Te + 6OH^- \quad \varphi^\theta = -0.57\ V$$

   $$N_2H_4 + 4OH^- \rightarrow N_2 \uparrow + 4H_2O + 4e^- \quad \varphi^\theta = -1.16\ V$$

   where $\varphi^\theta$ is the standard electrode potential. They are measured at 298 K and under standard pressure ($p^\theta = 100 kPa$). According to Nernst equation, $\Delta G = -zEF$, where z is the number of transferred electrons, E is the electromotive force, and F is the Faraday constant, which is 96500 C·mol$^{-1}$, the overall reaction is spontaneous change because of the negative change of Gibbs free energy, which seems not to be related to the temperature. However, the half reaction of hydrazine oxidation is endothermic reactions, driven by the increase of entropy. The higher temperature promotes the forward reaction rate in this half reaction, leading to higher productivity of 2D Te. It can be seen in Supplementary Fig. 8 that the productivity dramatically increased from 160 °C to 180 °C. But there is no significant difference between 180 °C and 200 °C, possibly due to the breaking of the weak van der Waals bonds between Te chains and the damaging the 2D Te nanostructures by the high temperature with extra energy of Te atoms. There may be a balance between the improvement of the degree of reaction and the excessive energy decomposing the nanostructure. These results warrant further in-depth investigations.

   **Thickness-mobility dependence fitting**

   The field-effect mobility displays a non-monotonic dependence on Te film thickness. Thomas-Fermi screening effect[10] has been successfully applied to model the total conductance in a biased 2D material film with a certain thickness, such as graphene, MoS$_2$, and black phosphorus[11-13]. Here we adopted the same method to fit the mobility-thickness relationship. Due to charge screening, the mobility *μ* and carrier density *n* in thin films are no longer uniform but a function of the depth from the interface *z*. Considering a slab with infinitesimal thickness *dz* at depth *z*, the total conductance σ(z+dz) will be the conductance of this thin slab plus the total conductance of σ(z) in series with two interlayer resistance $R_{int}$ to form a resistor network as shown in Fig. S15, which gives recursion equation[11]:

   $$\sigma(z + dz) = qn(z)\mu(z)dz + \frac{\sigma(z)\frac{1}{2R_{int}}}{\sigma(z) + \frac{1}{2R_{int}}} \quad (1)$$

   By replacing $R_{int}$ with the resistance of unit length *r* multiplied by *dz*, we can reform the eq. (1) into a differential equation:



$$\frac{d\sigma}{dz} = qn(z)\mu(z) - \sigma(z)^2 r \qquad (2)$$

The Thomas-Fermi screening effect goes that the carrier density and mobility decays as depth increases with a characteristic screening length $\lambda$. Therefore we can express n(z) and µ(z) as[12]:

$$n(z) = n(0)\exp\left(-\frac{z}{\lambda}\right) \qquad (3)$$

$$\mu(z) = \mu_{inf} - (\mu_{inf} - \mu_0)\exp(-\frac{z-z_0}{\lambda}) \qquad (4)$$

where $\mu_{inf}$ and $\mu_0$ are the mobility at infinity and the interface respectively.

By substituting eq.3 and eq.4 back to eq.2, we can derive an expression for total conductance σ(z). Finally, we convert the total conductance into effective mobility with the simple relation:

$$\mu_{FE}(z) = \frac{\sigma(z)}{Q_{tot}} \qquad (5)$$

where $Q_{tot}$ is the total gate-induced charge in the entire channel which can be estimated by multiplying gate voltage by $C_{ox}$.

This model fits well with our experiment data, and the calculated Thomas-Fermi screen length is 4.8 nm which is a reasonable value compared to other material systems such as $MoS_2$ and black phosphorus.

**Thickness-dependent on/off ratio in 2D Te FETs**

Most of the 2D FETs are operated in accumulation-mode or depletion-mode junction-less type field-effect transistors with oxide as the dielectric. Their device operation is very similar to III-V MESFETs or HEMTs and thin film transistors such as Indium–gallium–zinc oxide (IGZO) ones. They are very different from the conventional Si MOSFETs which are operated in inversion-mode and independent on film thickness. A simplified model to demonstrate the degradation of on/off ratio with thickness is presented following. At on-state, the carrier transport is mainly contributed by gate induced accumulated carriers located within a few nanometers from the interface and the doped channel. Therefore, the on-state current varies several times for a wide range of thickness as confirmed by the experimental data. We define the off-state to be the scenario where the Fermi energy is at charge neutral point at the interface, as shown in Figure 1 (for simplicity, we ignore Schottky contact impact and electron and hole mobility difference in off-state). The surface potential or band bending at off-state is deduced to be around 0.15 V, which is a reasonable estimation considering the 0.35 eV bandgap of Te and the fact that the Fermi level of bulk Te is closer to its valence band. Under depletion approximation, we can then derive the surface potential at certain depth x (distance to the interface) by solving Poisson's equation:



$$\varphi_s(x) = \frac{qN_A}{2\varepsilon_s\varepsilon_0}(x - x_d)^2,$$

where $\varepsilon_s$ is the permittivity of Te, $\varepsilon_0$ is the vacuum permittivity, $N_A$ is the intrinsic doping level and $x_d$ is the maximum depletion width which is calculated to be

$$x_d = \sqrt{2\varepsilon_s\varepsilon_0\varphi_s(0)/qN_A} \approx 22\ nm,$$

close to experimentally observed thickness range where the on/off ratio starts to saturate. The hole density at $x$ then can be expressed by:

$$p(x) = p_0 \exp(-q\varphi_s(x)/kT).$$

We plotted the surface potential and carrier distribution as a function of distance $x$ in Figure 2. By integrating carrier density $p(x)$ with $x$, the off-state sheet carrier density can be numerically calculated as a function of flake thickness t:

$$n_{2D\_off}(t) = \int_0^t p(x)dx,$$

as shown in Figure 3. We can see that the off-state carrier sheet density increases by more than 4 orders as the thickness increases from monolayer to over 22 nm. Since

$$I_{on}/I_{off} \approx 1/I_{off} \propto 1/n_{2D\_off}(t),$$

we expect the on/off ratio degrades as carrier density increases in thicker flakes. If the film thickness is larger than the maximum depletion width, the device cannot be turned off, and the situation becomes trivial and obvious.

Noted that the above discussion is barely a simplified illustration why on/off ratio has such a thickness-dependent trend whereas in real devices the situation can be much more complicated.

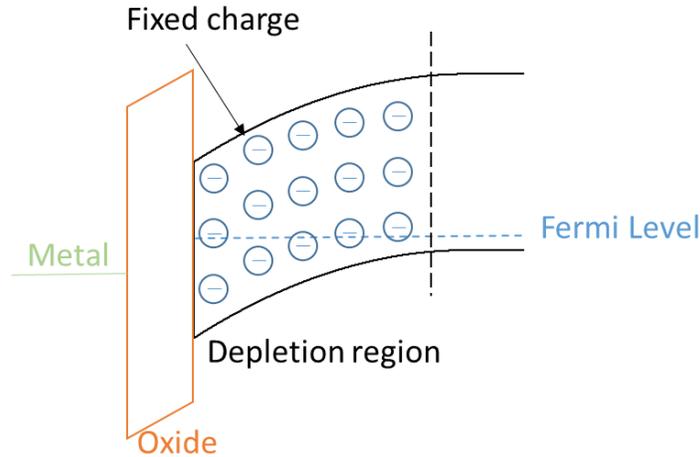

*Figure 1. Band diagram at along MOS stack direction. We define the off-state to be the scenario where the Fermi level at the interface is located at the charge neutral level.*



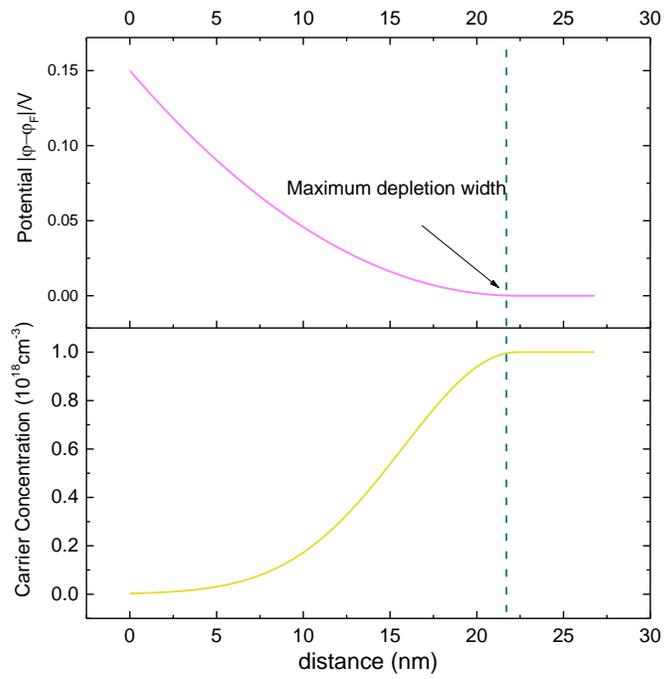

***Figure 2.*** *The potential (upper) and carrier concentration (lower) distribution at off-state as a function of distance from the semiconductor-oxide interface. The maximum depletion width is around 22.3 nm.*

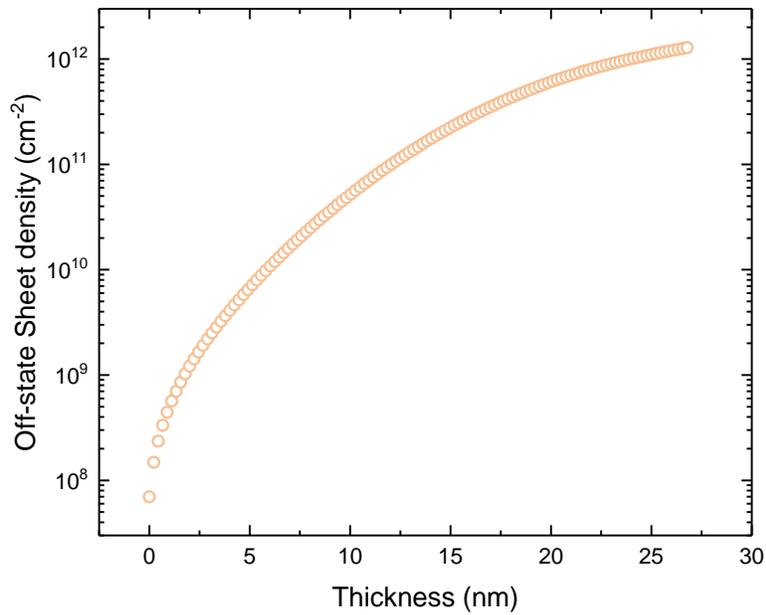

***Figure 3.*** *The 2D sheet density at off-states for flakes with different thickness.*

**Supplementary Figures**

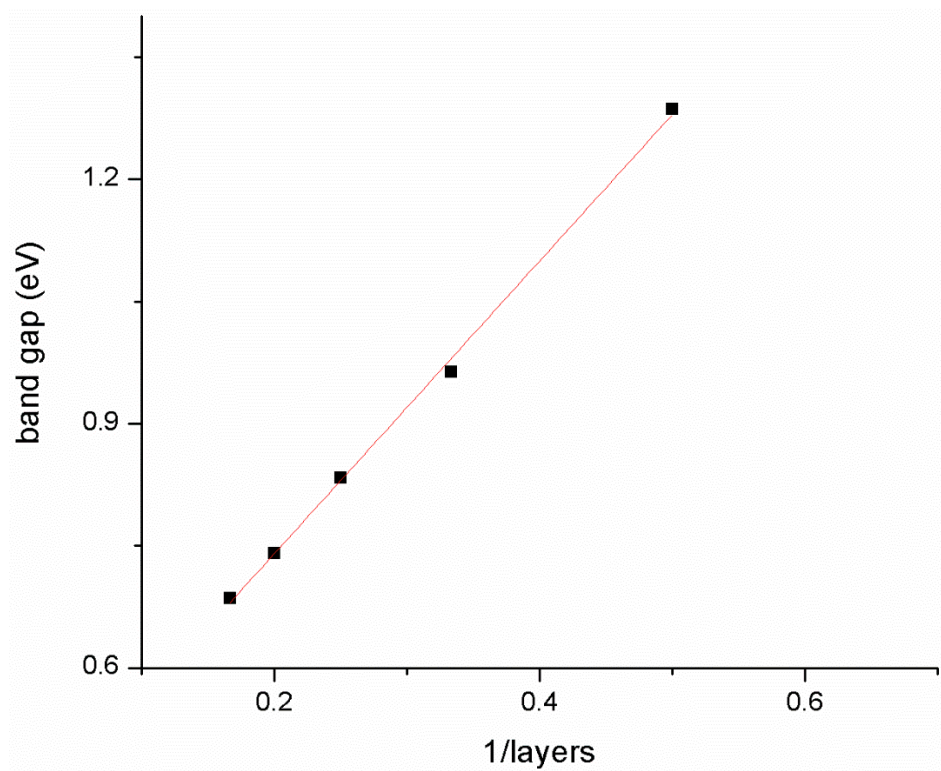

**Supplementary Figure 1** | Thickness-dependent bandgap for tellurene, calculated by HSE functional. The bandgap shows a linear dependence on the inverse number of layers, following $E_g = 0.38 + 1.8/n$ (eV). As n goes to infinity, this relation gives a band gap of 0.38 eV for the bulk Te, in good agreement with the experimental data. Given the interlayer distance ~ 3.91 Å, this relation can be rewritten as $E_g = 0.38 + 1.8*3.91/t$ (eV), where t is the thickness (Å).



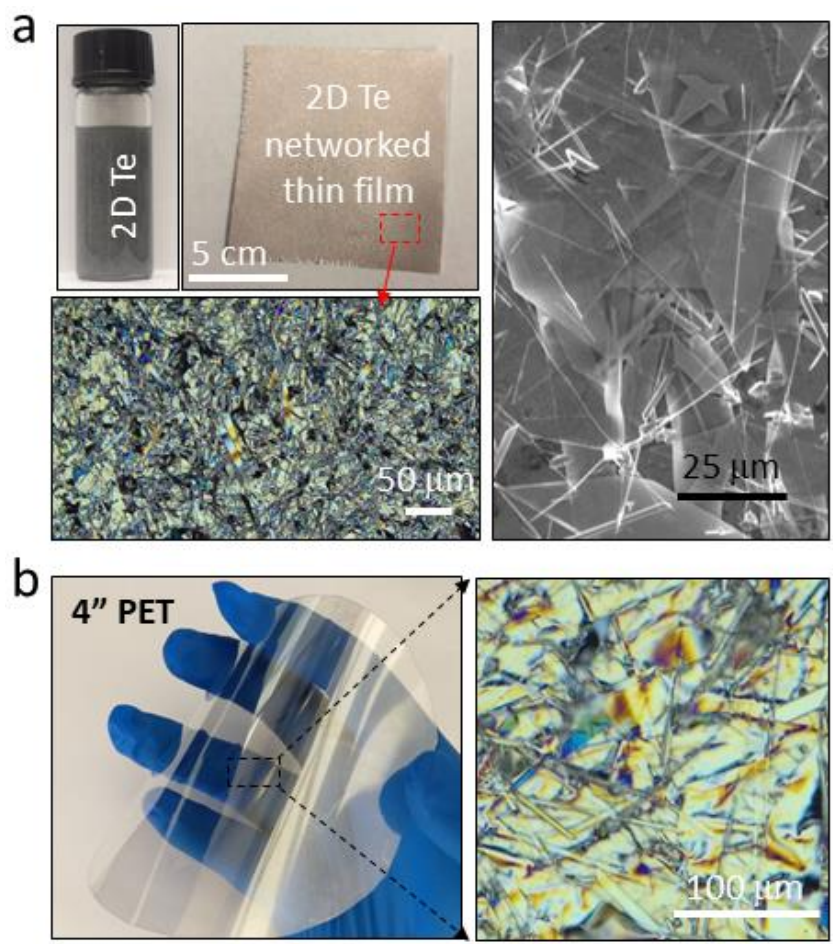

**Supplementary Figure 2** | Large-scale transfer and assembly of 2D tellurene into **(a)** networked thin film through ink-jet printing and **(b)** monolayer thin film through LB method.



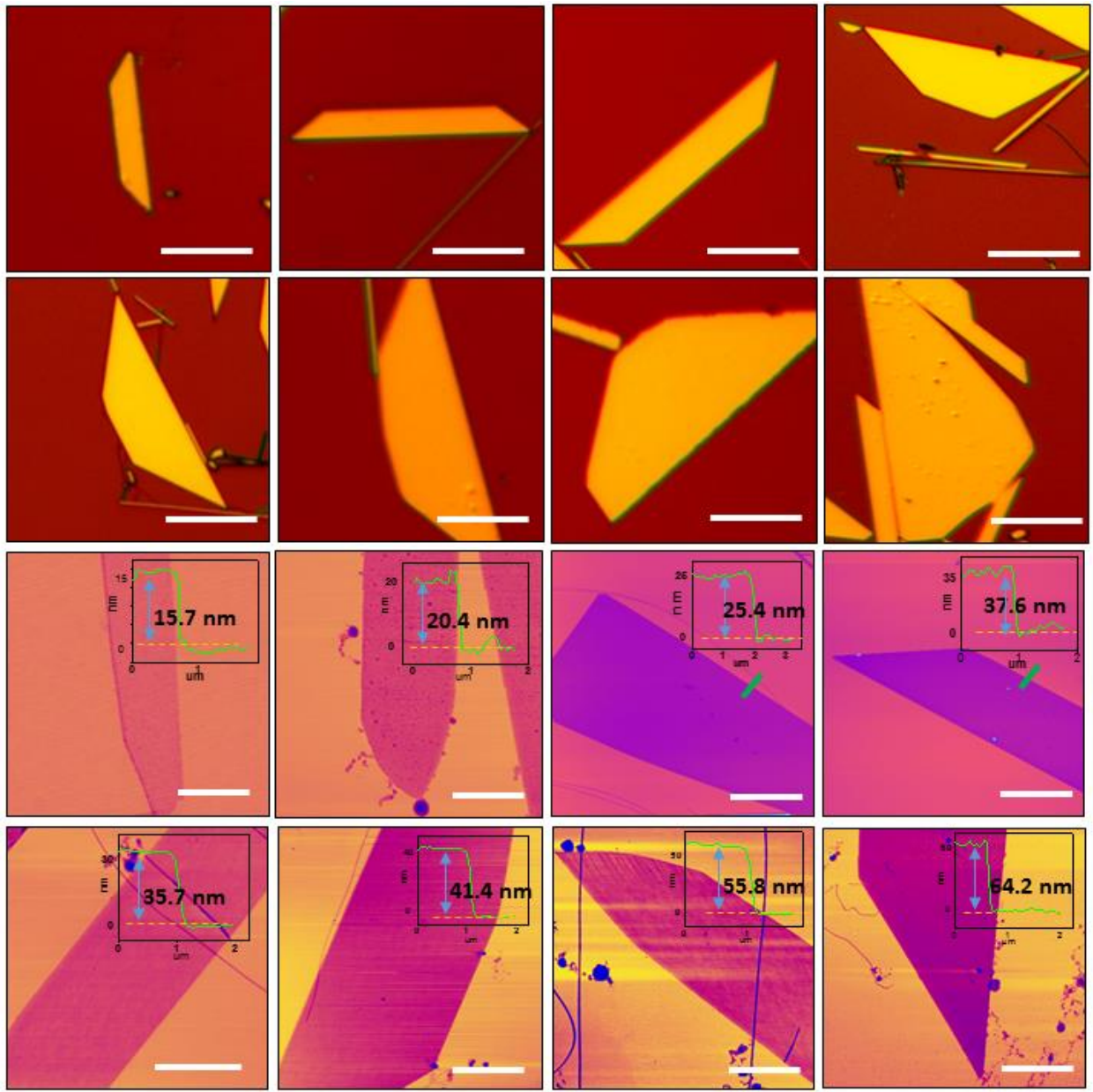

**Supplementary Figure 3** | Optical and AFM images of tellurene flakes with various edge lengths and thicknesses. The scale bar is 20 μm.



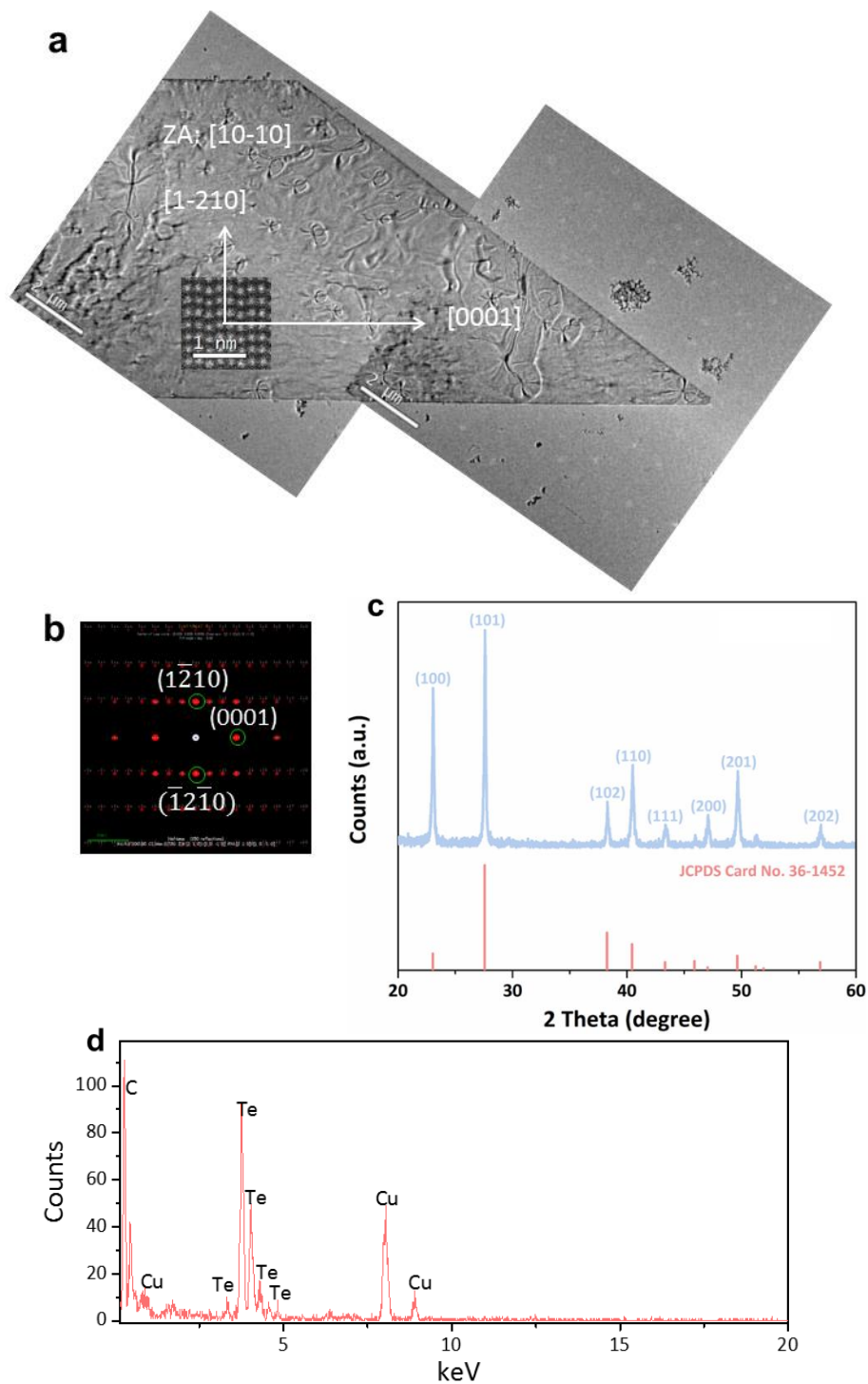

**Supplementary Figure 4** | Structural, composition, and quality characterization of tellurene. **a,** Low-resolution TEM image of a tellurene flake. The contour contrast is due to bending of the flake. **b,** Simulated diffraction pattern. **c,** XRD result of 2D Te crystals show high crystallization without impurity peaks in the material, in good agreement with JCPDF card number:36-1452 for tellurium. **d,** EDS spectra confirming the chemical composition of Te. The Cu peaks in EDS spectra come from Cu TEM grid.



| Time (hrs) Na₂TeO₃/ PVP | 1.5 | 2 | 3 | 5 | 7 |
|---|---|---|---|---|---|
| 261.8 | 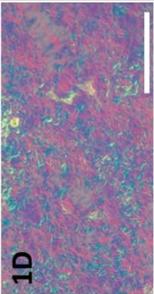 1D | 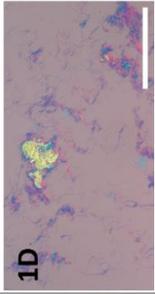 2D | 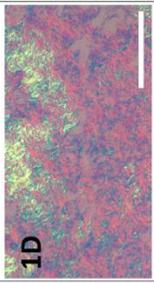 2D | 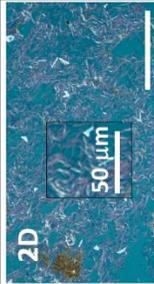 2D | 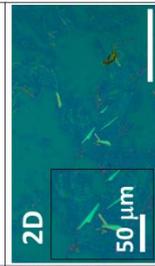 2D |
| 87.3 | 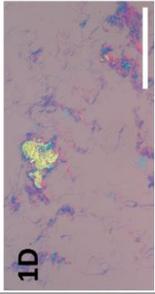 1D | 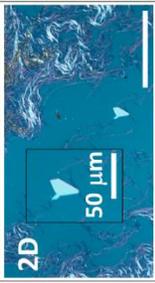 1D | 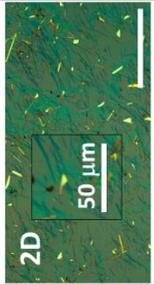 2D | 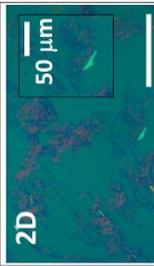 2D | 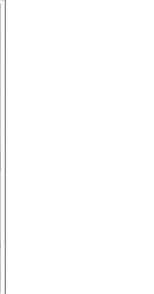 2D |
| 52.4 | 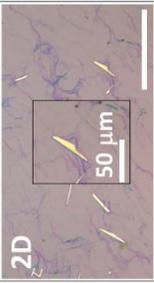 1D | 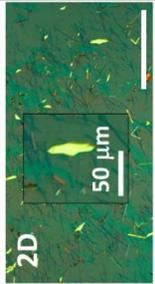 1D | 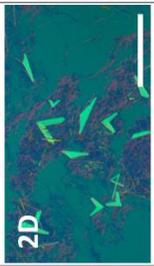 1D | 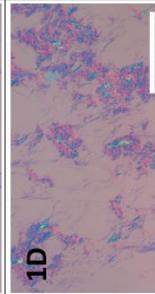 2D | 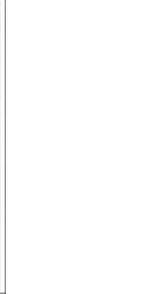 2D |
| 37.4 | 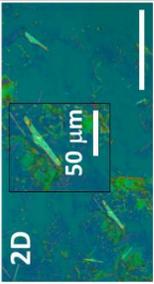 1D | 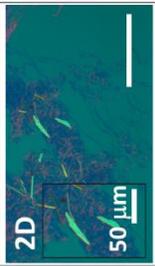 1D | 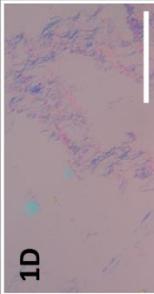 1D | 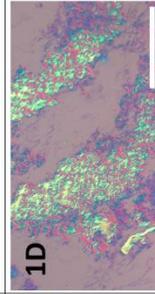 2D | 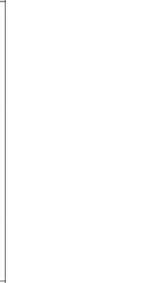 2D |
| 29.1 | 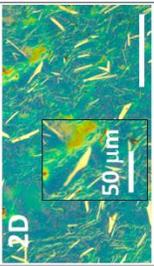 1D | 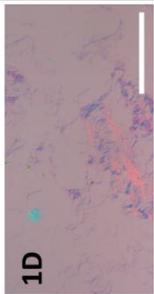 1D | 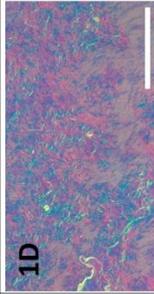 1D | 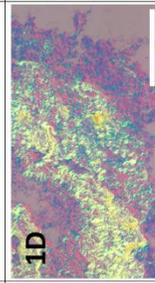 1D | 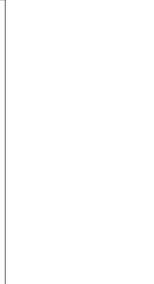 2D |



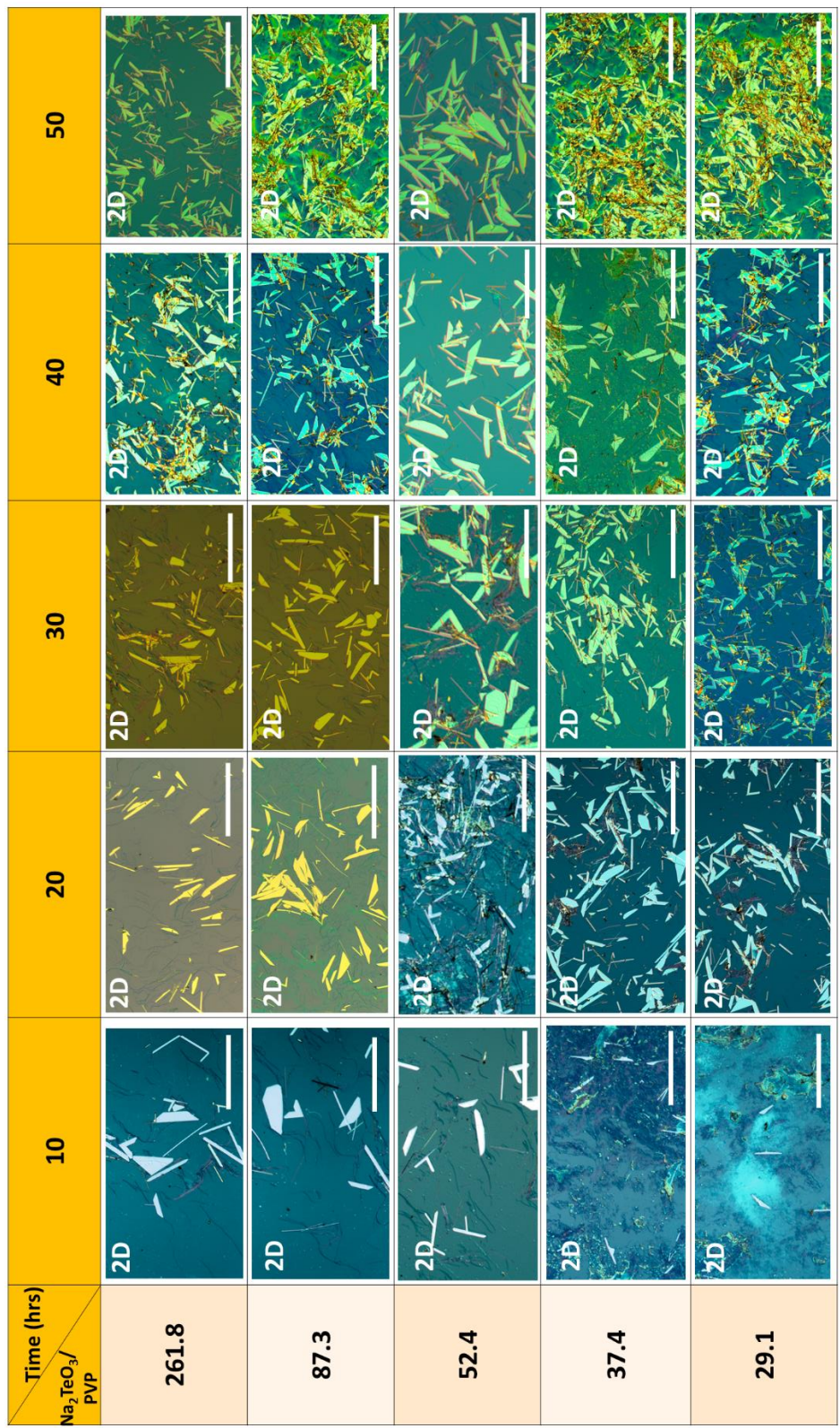

**Supplementary Figure 5 | Morphology evolution in the growth from early to final stages.** The scale bars are 200 µm and 50 µm in the main figures and insets.



a

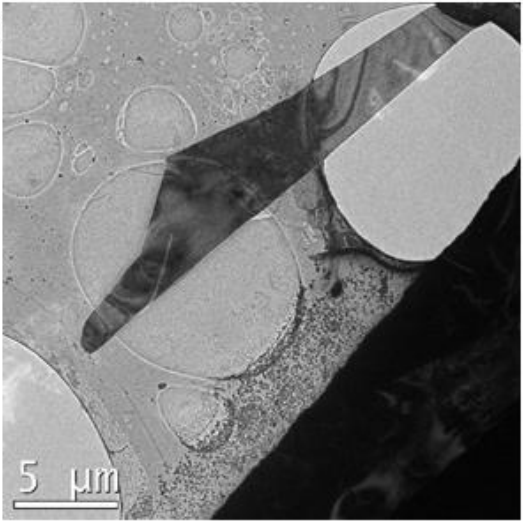
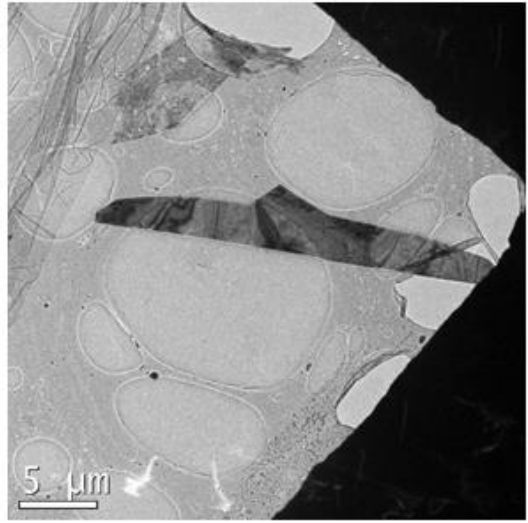
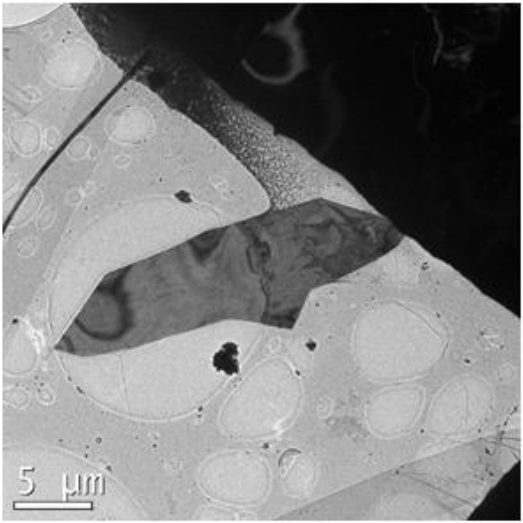
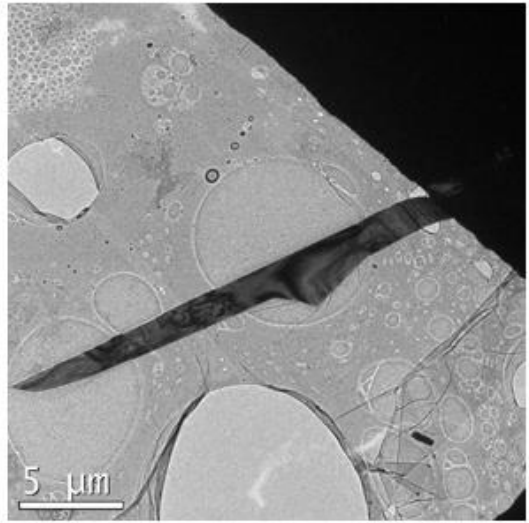



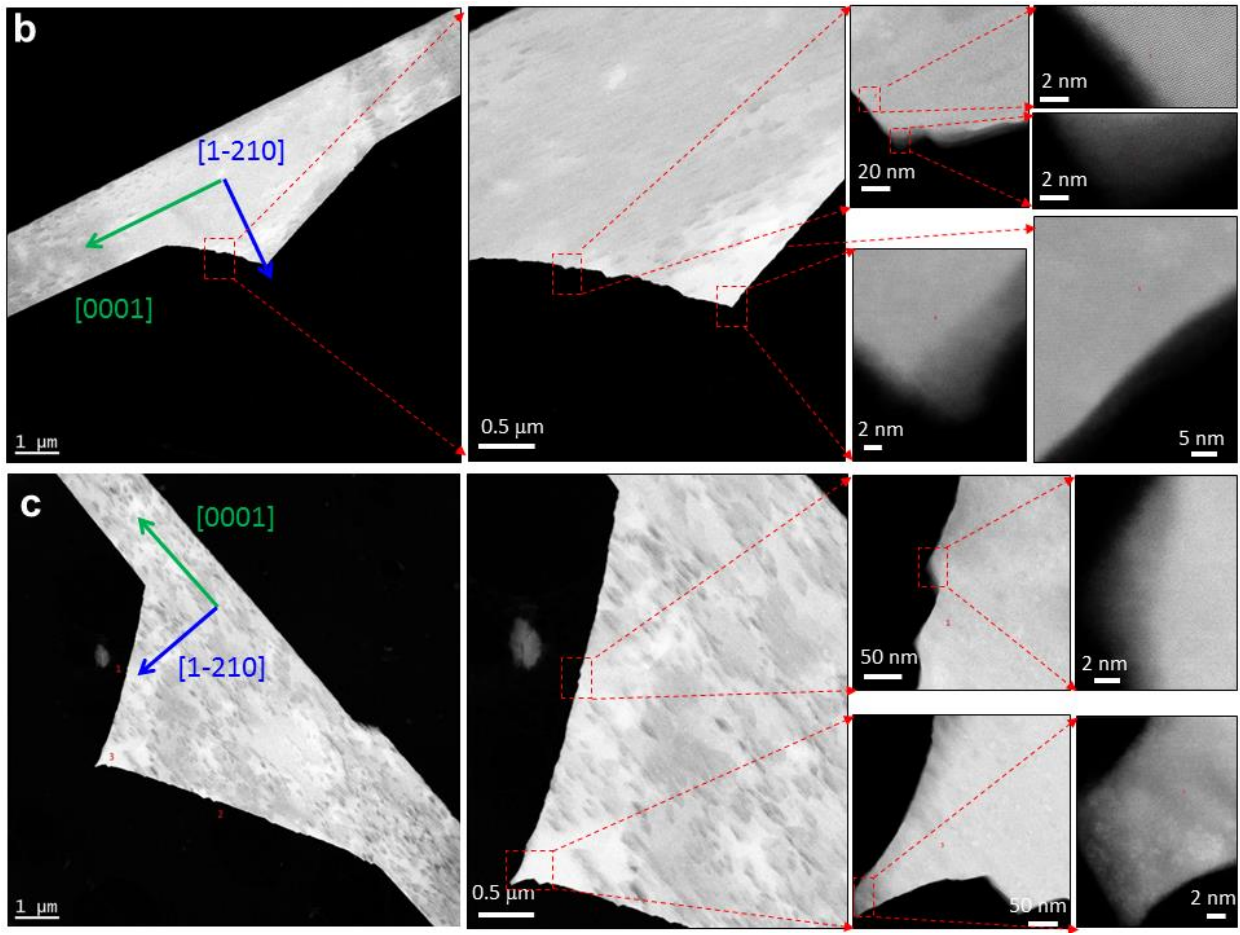

**Supplementary Figure 6 | a,** Low-resolution TEM images of the intermediate structures with edges showing different angles with respect to [0001] direction. **b** and **c**, STEM images of two such structures showing that the edges are made of many steps.



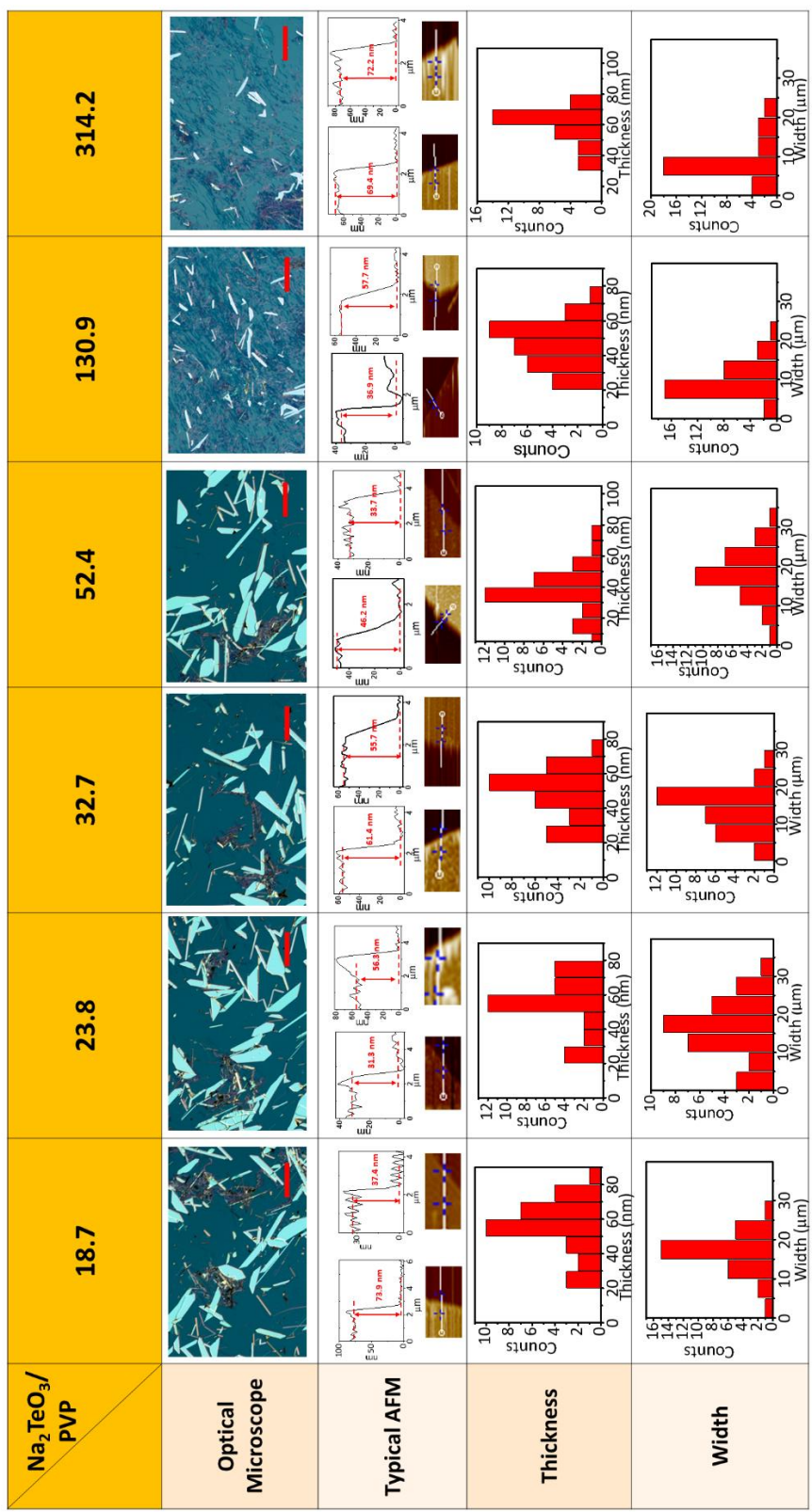

| Group | Na$_2$TeO$_3$/PVP |
|---|---|
| 1 | 16.4 |
| 2 | 17.5 |
| 3 | 18.7 |
| 4 | 20.1 |
| 5 | 21.8 |
| 6 | 23.8 |
| 7 | 26.2 |
| 8 | 29.1 |
| 9 | 32.7 |
| 10 | 37.4 |
| 11 | 43.6 |
| 12 | 52.4 |
| 13 | 65.5 |
| 14 | 87.3 |
| 15 | 130.9 |
| 16 | 261.8 |
| 17 | 288.0 |
| 18 | 314.2 |
| 19 | 340.3 |

**Supplementary Figure 7** | AFM data supporting Fig. 2c thickness. The width is the direction which is vertical to <0001>. The scale bar is 50 μm. The table shows the detailed PVP concentration for each group in Fig. 2c.



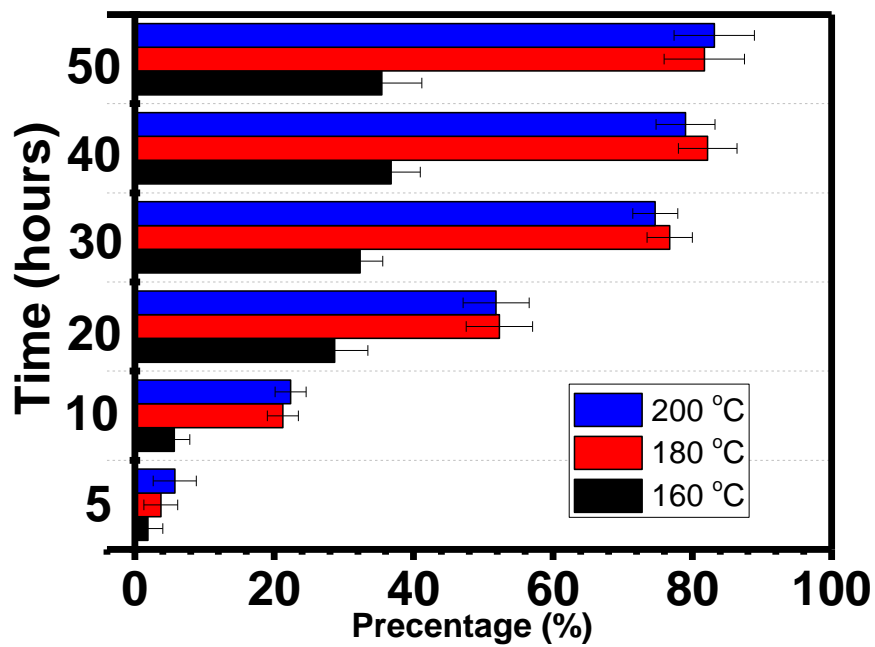

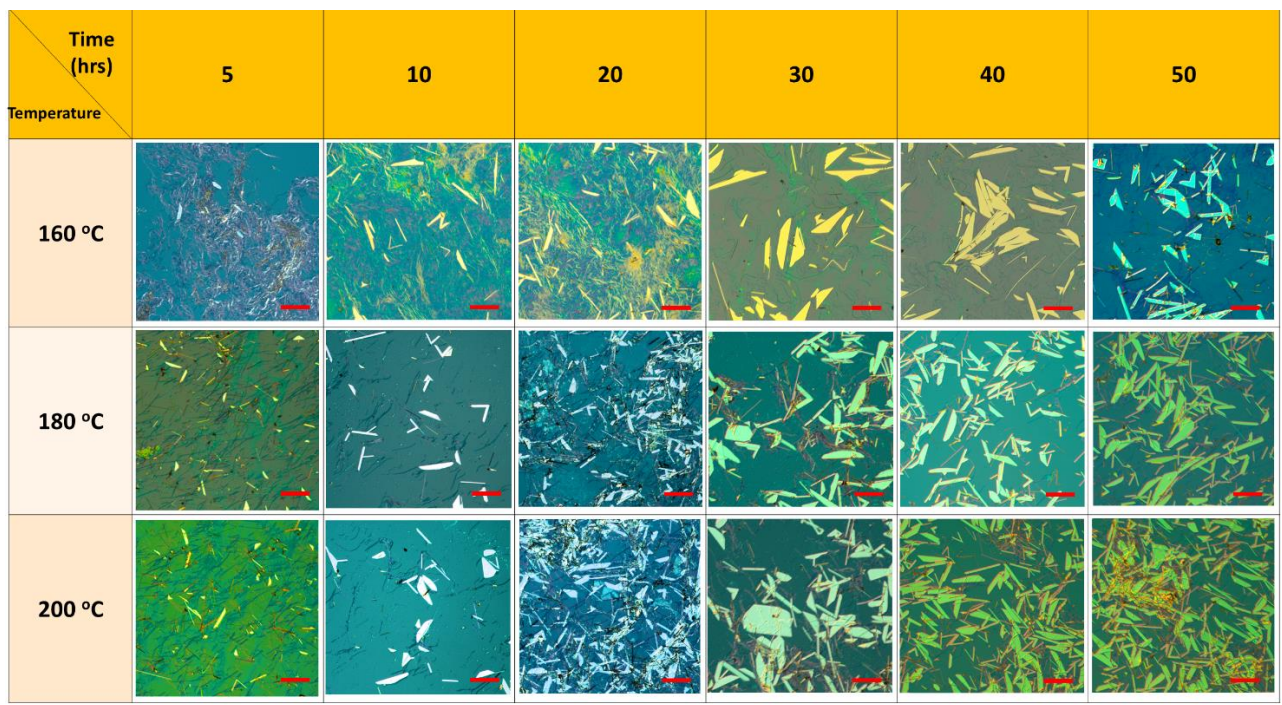

**Supplementary Figure 8** | Productivity of tellurene grown at different reaction temperatures. The Na$_2$TeO$_3$/PVP ratio is 52.4/1 for all reactions here. Scale bar: 50 μm. Mean values from 5 technical replicates are indicated. Error bars represent s.d.



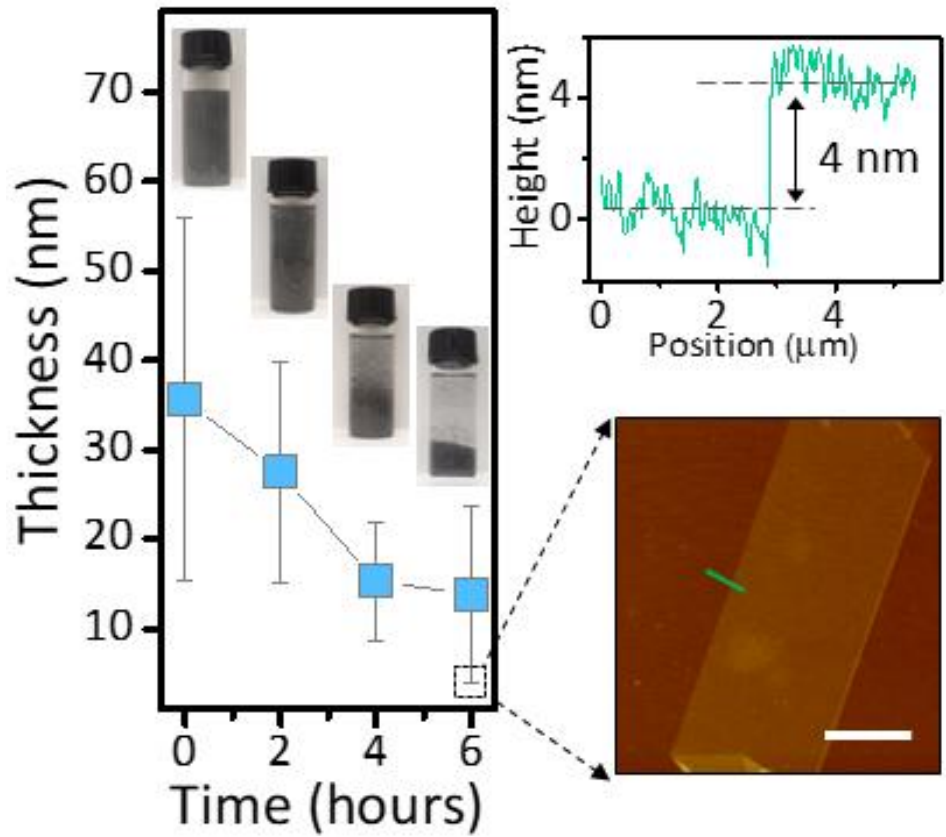

**Supplementary Figure 9** | Post-growth thinning process in the alkaline growth solution (pH ~ 11.5) for obtaining ultrathin few-layer tellurene. The scale bar is 5 μm. Mean values from 8 technical replicates are indicated. Error bars represent s.d.



| Time (hrs) | 0 | 2 | 4 | 6 | 8 | 10 |
|---|---|---|---|---|---|---|
| Water | 33.7 nm / 46.2 nm | 37.6 nm / 34.4 nm | 37.1 nm / 49.2 nm | 33.1 nm / 36.2 nm | 44.6 nm / 50.9 nm | 46.8 nm / 31.5 nm |
| IPA | 33.7 nm / 46.2 nm | 21.2 nm / 24.6 nm | 22.4 nm / 20.8 nm | 16.3 nm / 16.9 nm | 17.9 nm / 15.8 nm | 11.7 nm / 8.1 nm |
| Acetone | 33.7 nm / 46.2 nm | 12.9 nm / 21.3 nm | 8.7 nm / 11.2 nm | 6.1 nm / 8.4 nm | 13.5 nm / 7.5 nm | 7.1 nm / 10.6 nm |



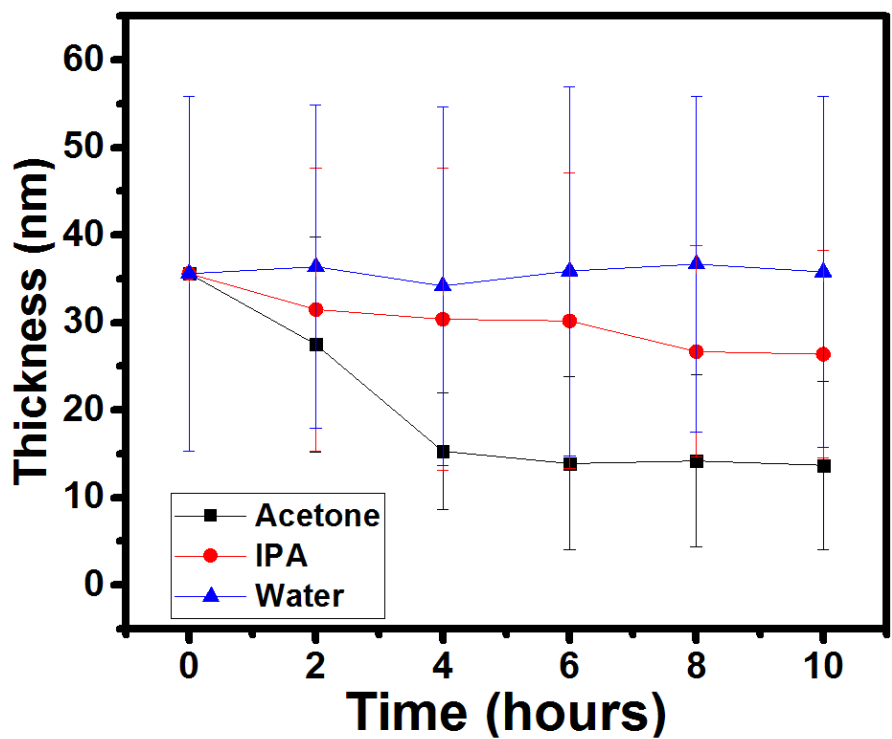

**Supplementary Figure 10** | Effect of different solvents on the post-growth thinning of tellurene in the alkaline growth solution (pH ~ 11.5). The scale bar is 50 μm. Mean values from 8 technical replicates are indicated. Error bars represent s.d.



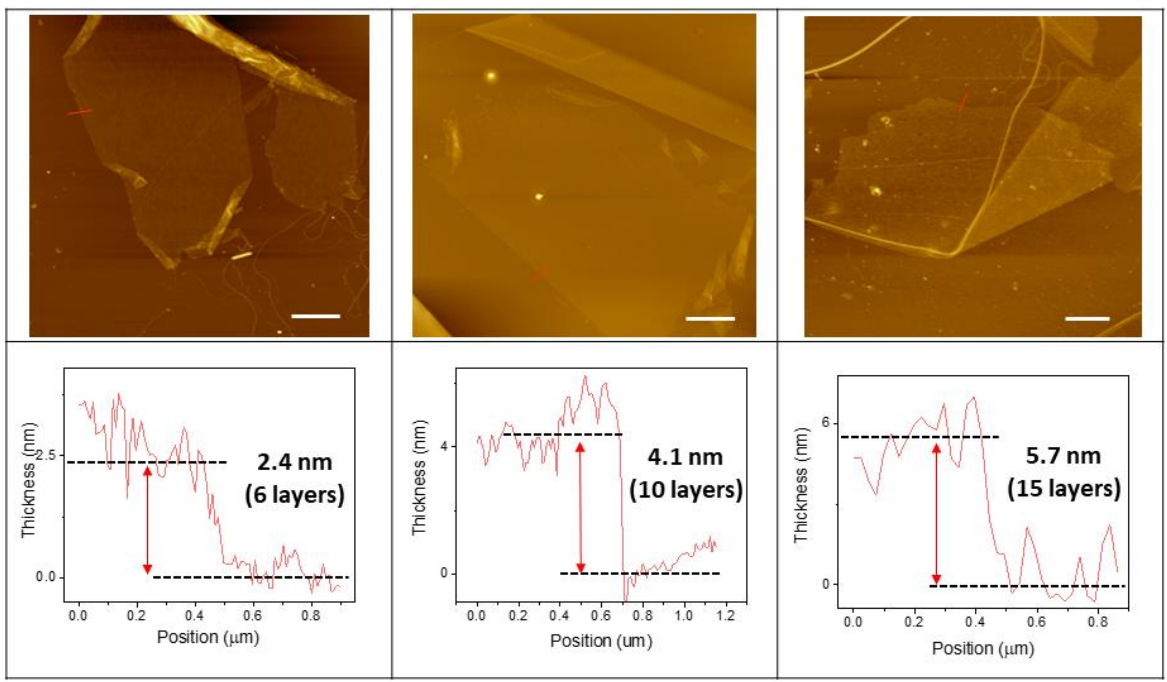

**Supplementary Figure 11** | Few-layer tellurene samples obtained after the post-growth thinning process (pH = 10.5) The scale bar is 5 μm.



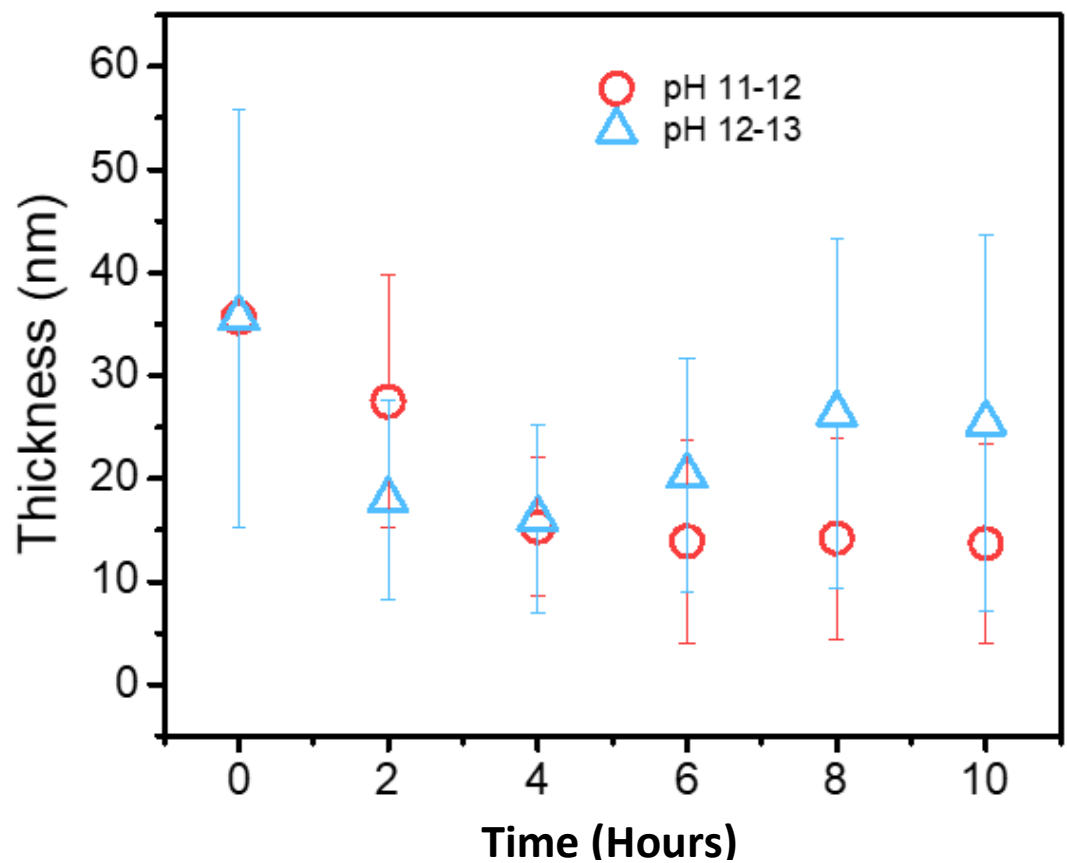

**Supplementary Figure 12** | Effect of pH values of the tellurene dispersion solution on the post-growth thinning processes. Mean values from 8 technical replicates are indicated. Error bars represent s.d.



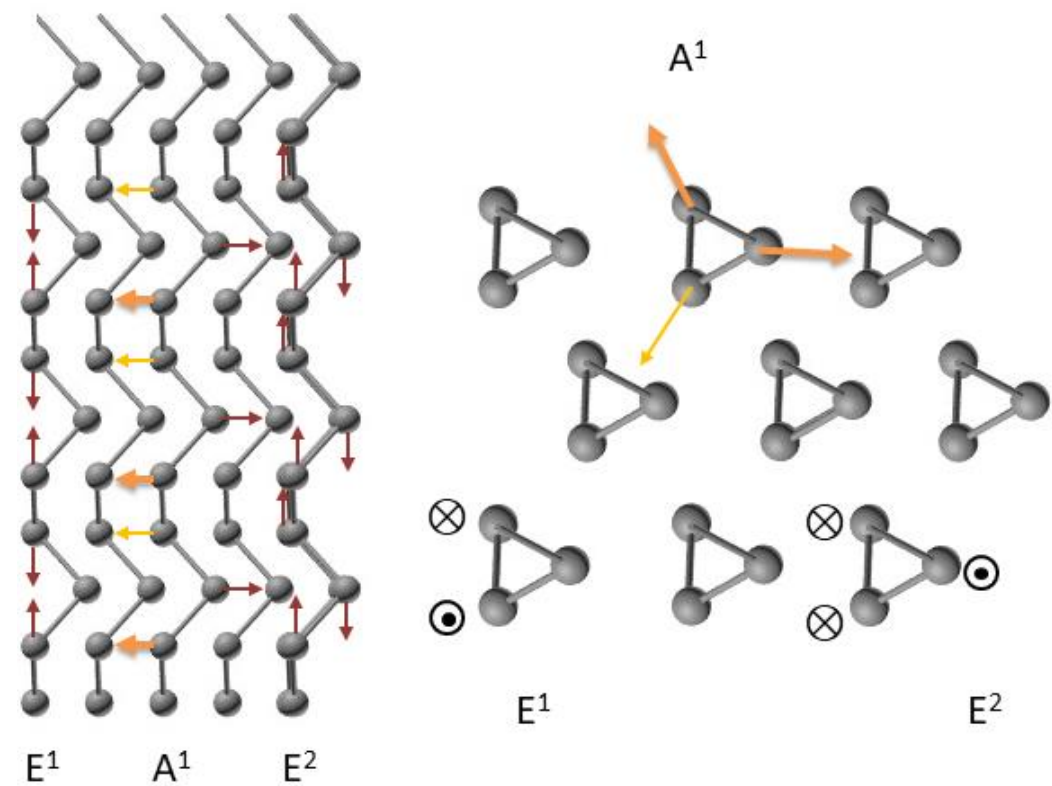

**Supplementary Figure 13** | Schematic shows the three main Raman-active modes in chiral-chain van der Waals material tellurene.



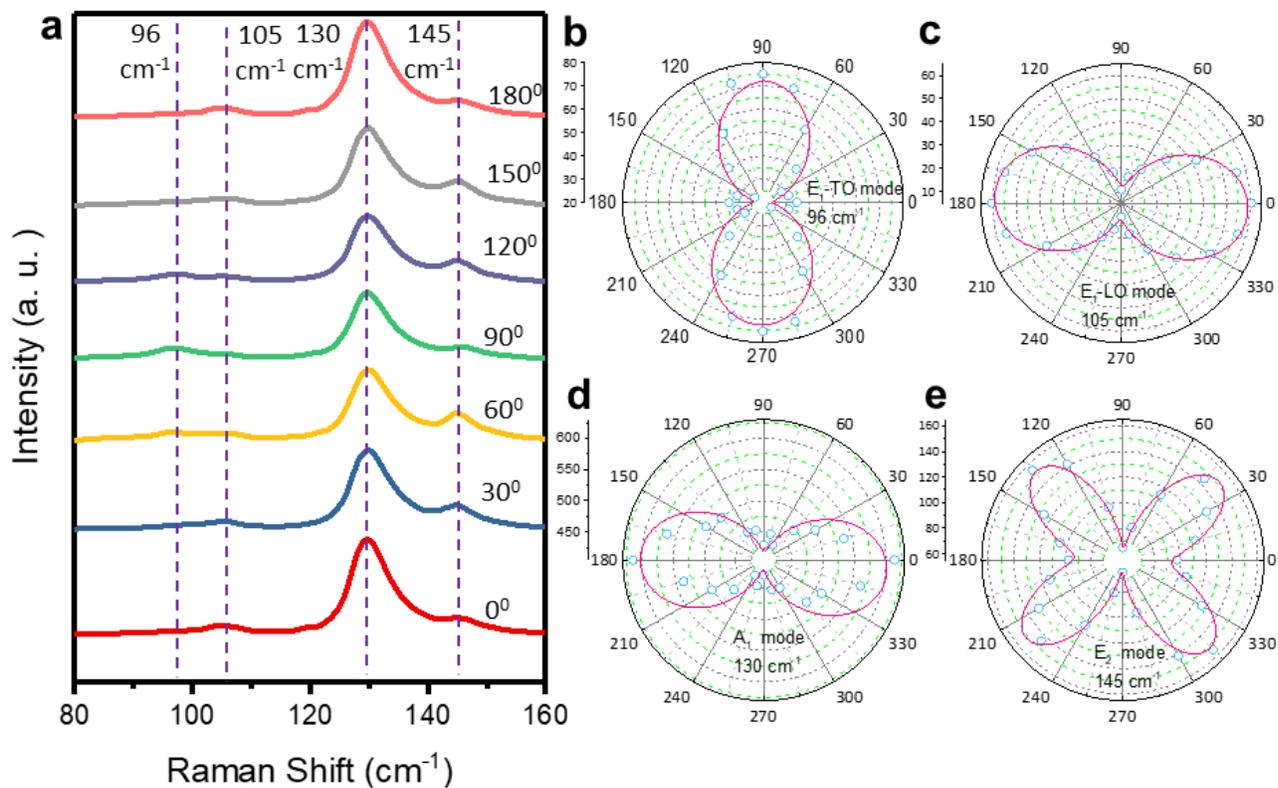

**Supplementary Figure 14 | Angle-resolved Raman Spectra for a 9.7 nm thick sample. a,** Raman spectra evolution with angles between crystal orientation and incident laser polarization. **b-e,** Polar figures of Raman Intensity corresponding to $A_1$ and two E modes located at 96 ($E_1$-TO), 105 ($E_1$-LO), 130 ($A_1$), and 145 ($E_2$) cm$^{-1}$. The fitting curves are described in Supplementary methods.



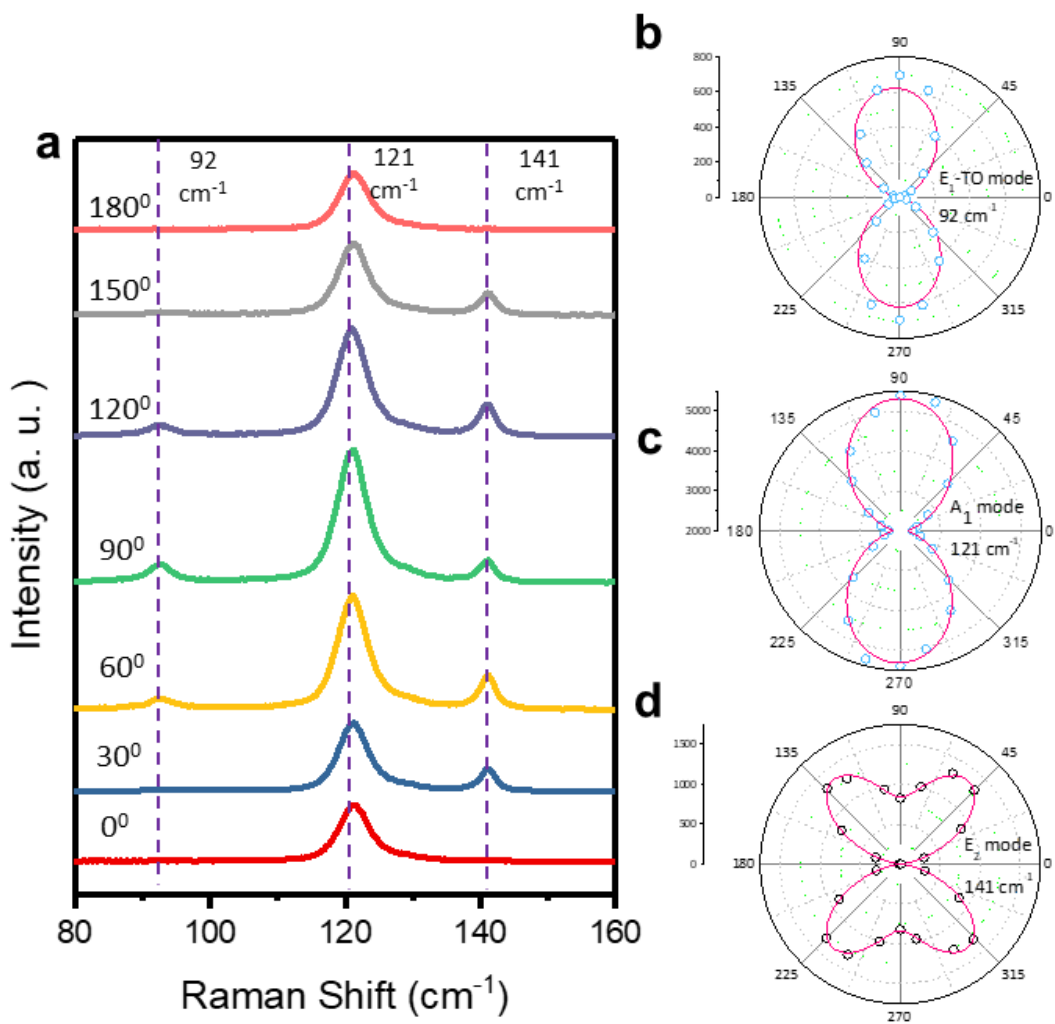

**Supplementary Figure 15 | Angle-resolved Raman Spectra for a 28.5 nm thick sample. a,** Raman spectra evolution with angles between crystal orientation and incident laser polarization. **b-d,** Polar figures of Raman Intensity corresponding to $A_1$ and two E modes located at 92 ($E_1$-TO), 121 ($A_1$), 141 ($E_2$) cm$^{-1}$. The fitting curves are described in Supplementary methods.



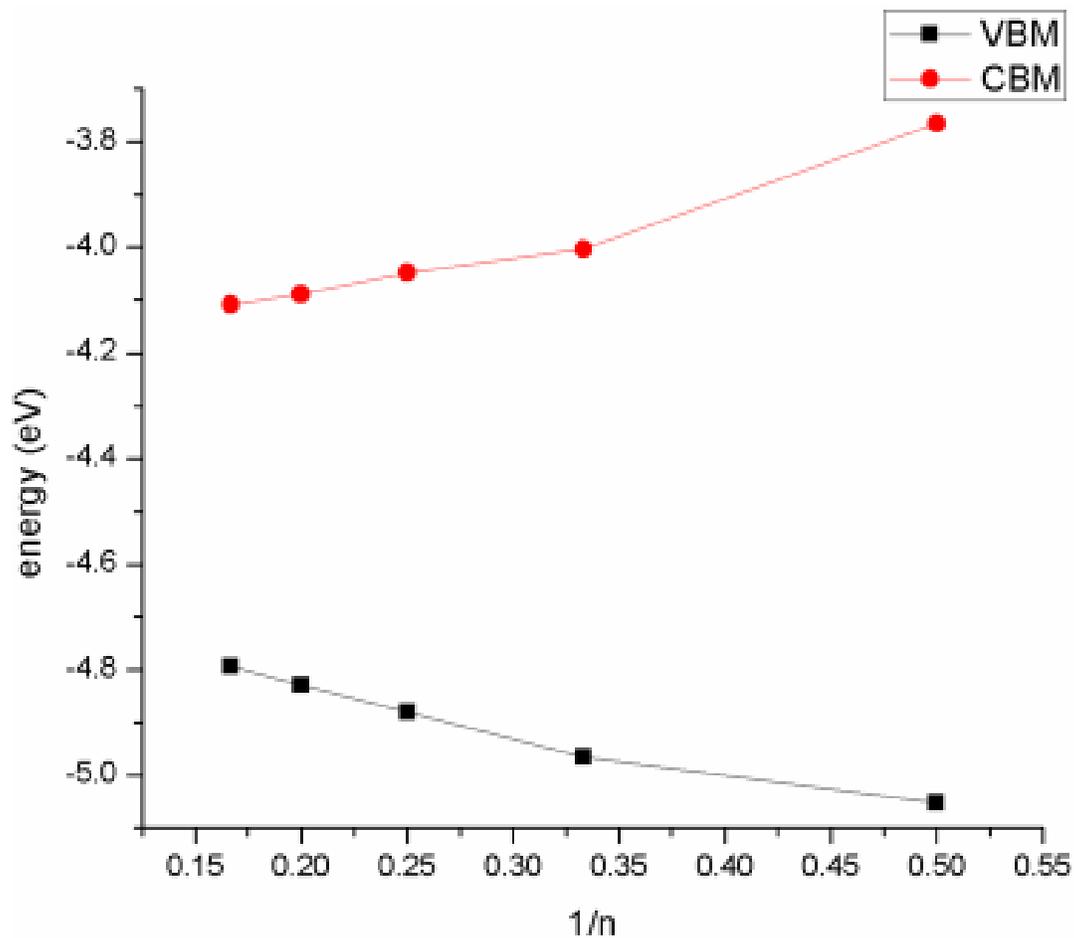

**Supplementary Figure 16** | Band edge level (with respect to vacuum) dependence on the number of tellurene layers, calculated by using HSE functional. In experiments, the thickness is > 6 layers. Therefore, the VBM should be at least higher than -4.8 eV. The Pd metal (used as a contact) has a work function of 5.22 - 5.60 eV, below the VBM of Te, so the Te shows a p-type behavior.



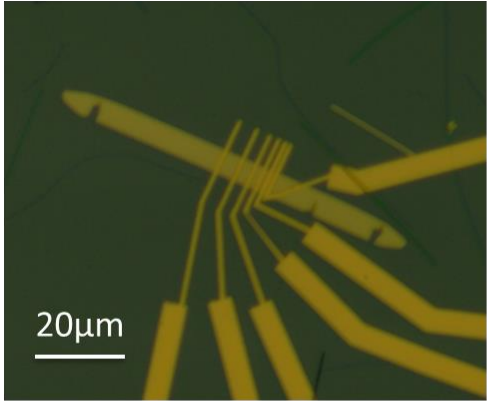
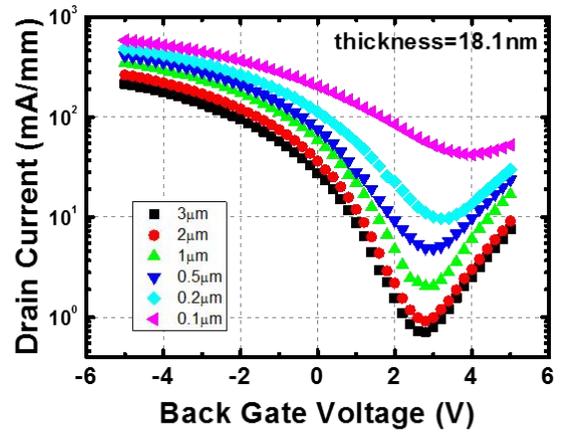
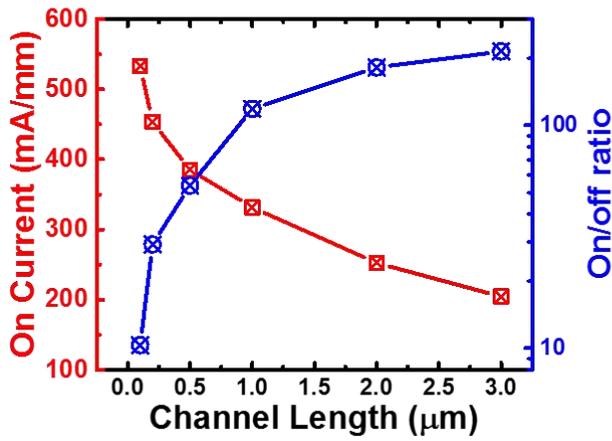
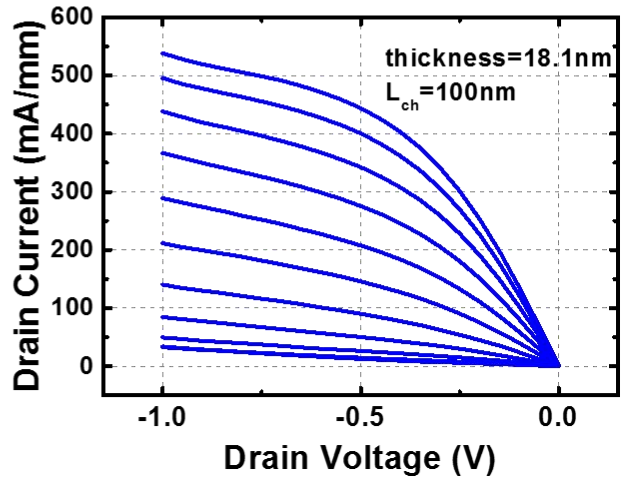

**Supplementary Figure 17** | By simply scaling down the channel length to 100 nm, the maximum on-current exceeds 550 mA/mm.



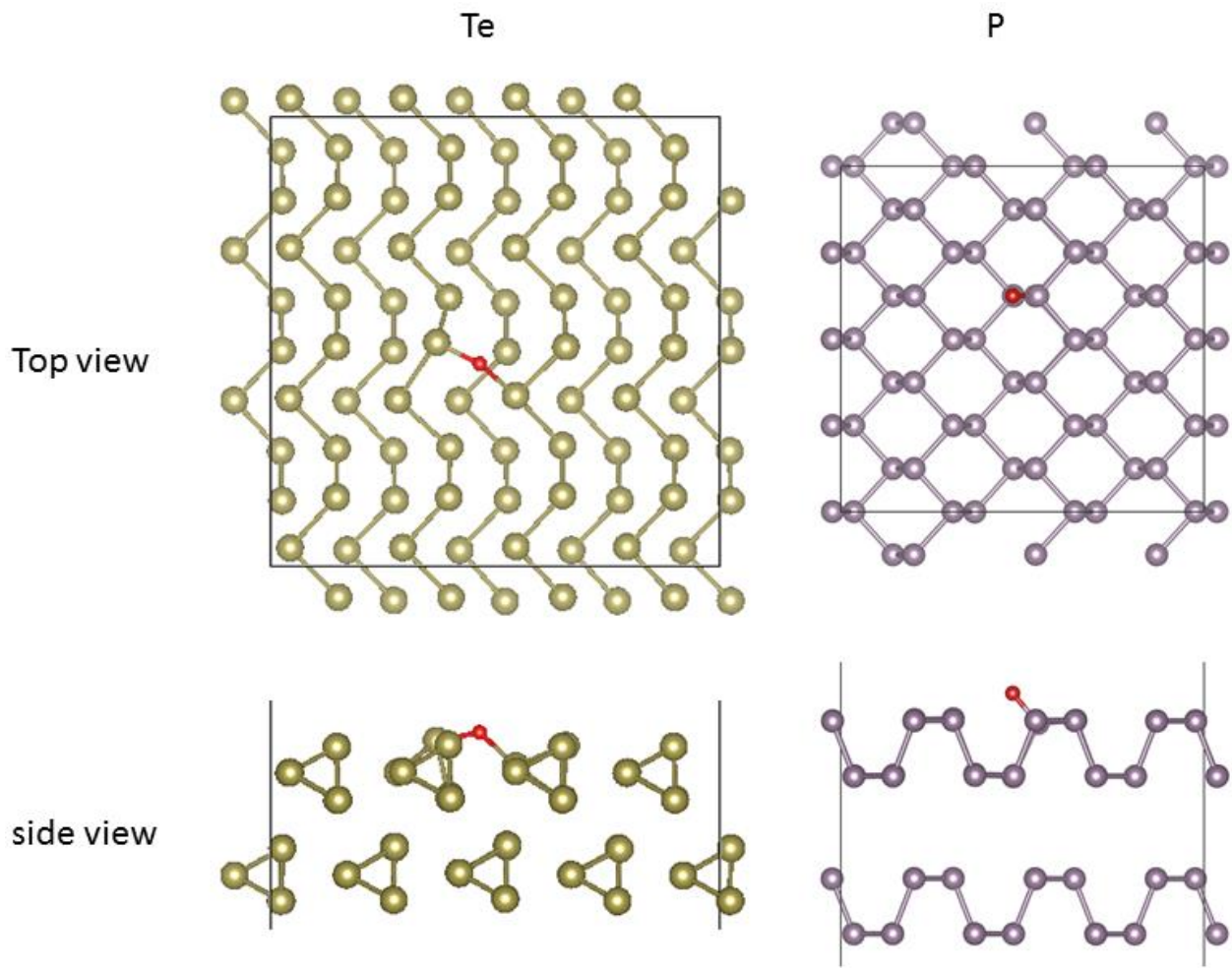

**Supplementary Figure 18** | Atomic structure of O adsorption on bilayer Te (left) and P (right). The black lines show the periodic boundaries. Note that the scale bars for Te and P are different. Compared with the binding in $O_2$, O atom binding to the surface is more favorable by ~-0.7 eV for bilayer Te, while ~ -2 eVs for bilayer P. This explains the superior stability of Te than black P.



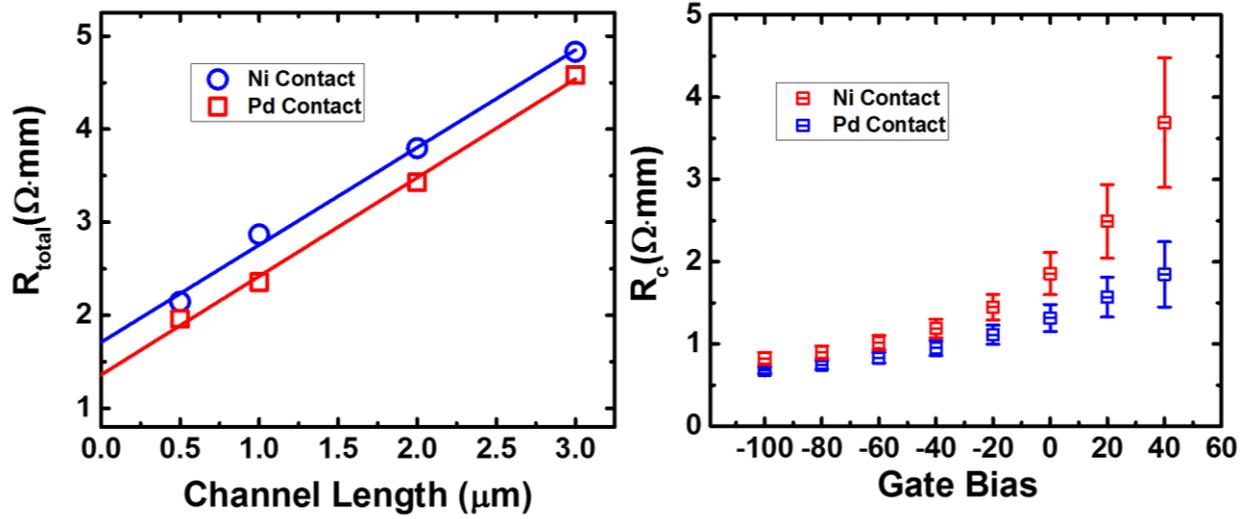

**Supplementary Figure 19** | Contact resistance was extracted by transfer length method. By extrapolating linear curve of total resistance vs. channel length, the contact resistance is extracted to be 0.67 and 0.82 Ω·mm on the same flake for Pd and Ni contacts respectively. Error bars arise from a standard error of linear fitting.



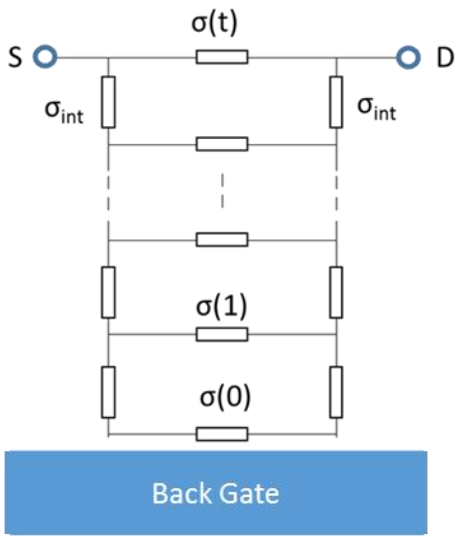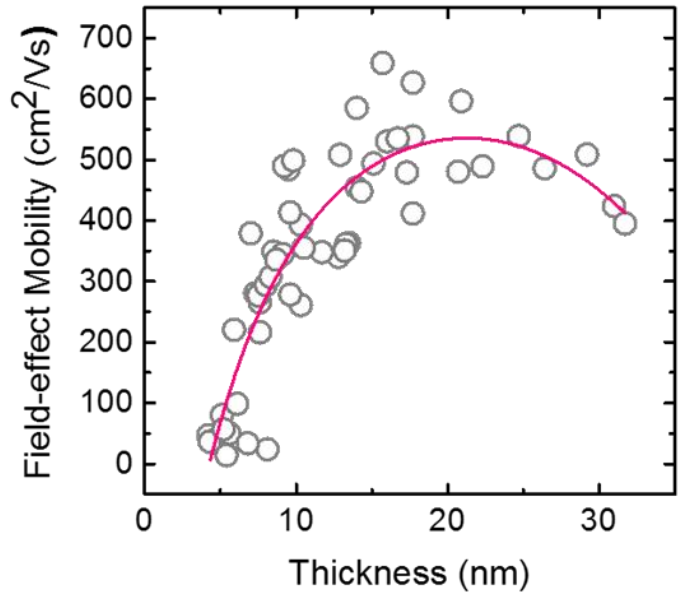

**Supplementary Figure 20 | Left,** the equivalent resistor network model describing total resistance with non-uniform conductance as a function of depth. **Right,** the fitting curve based on the Thomas-Fermi screening model.



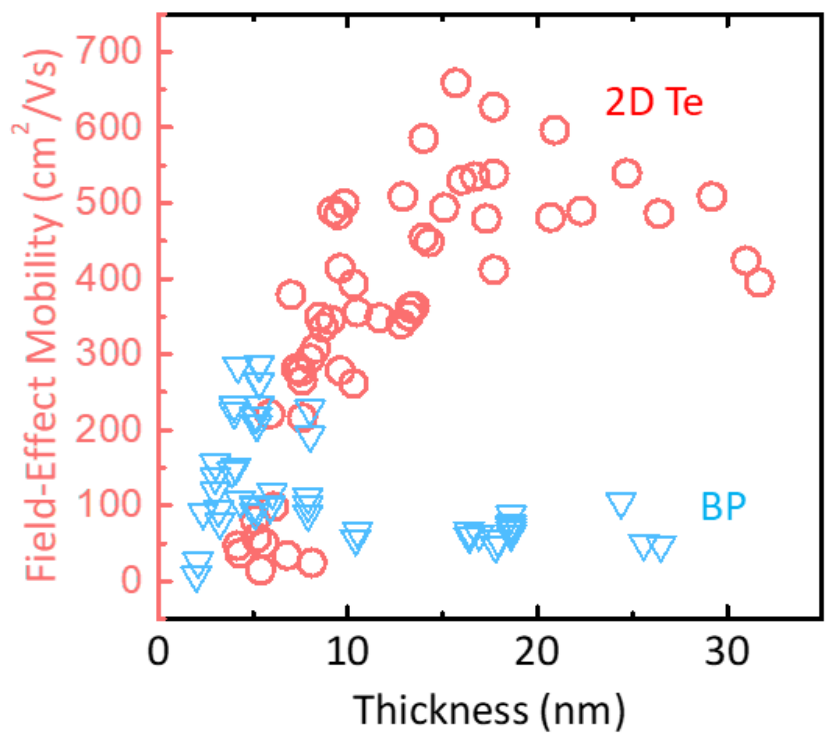

**Supplementary Figure 21** | Benchmark comparison of mobility between 2D Te and phosphorene using the same device structure, geometry, and mobility extraction method. The data for phosphorene is from Ref. 13.



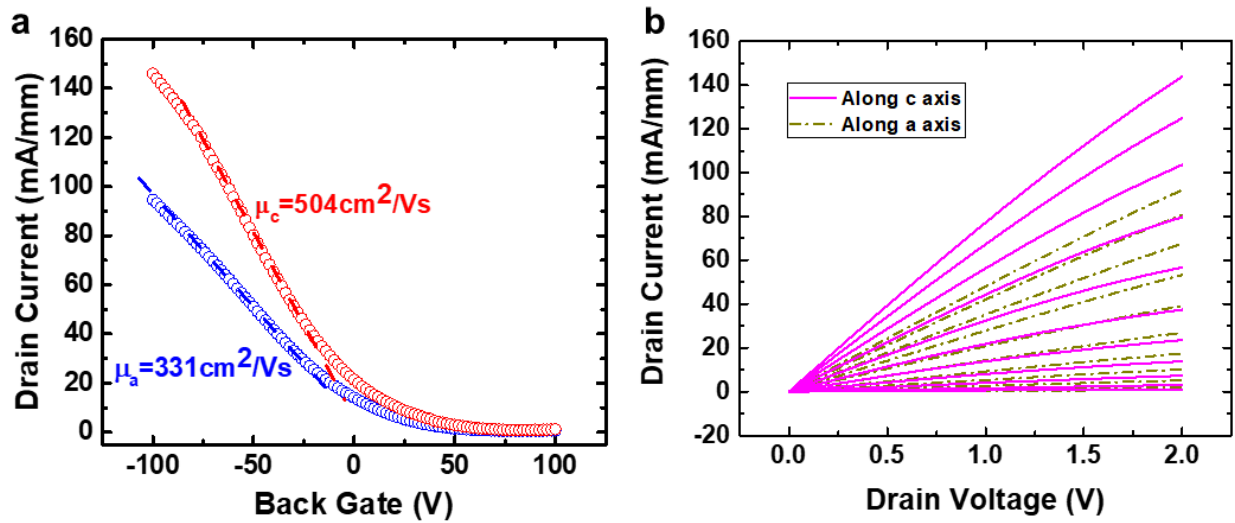

**Supplementary Figure 22** | The in-plane anisotropic electrical transport measurement for a 22-nm-thick sample. Here, the c-axis is the [0001] direction, and a-axis is the [1$\bar{2}$10] direction.



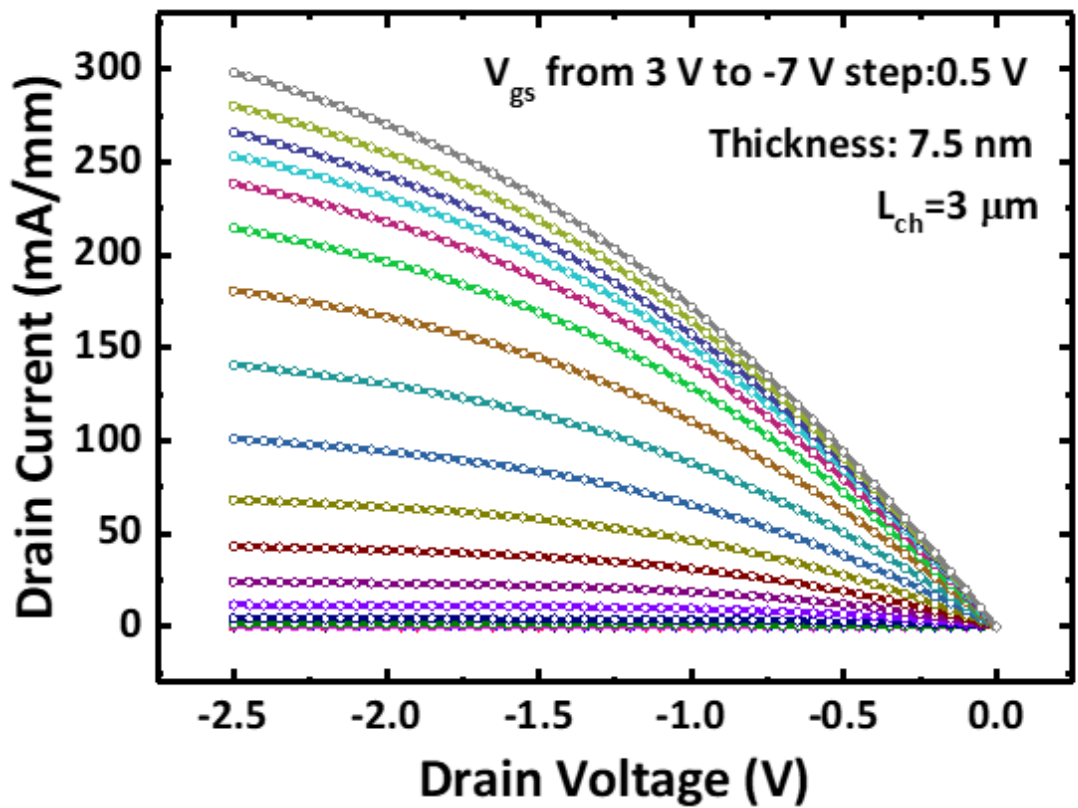

**Supplementary Figure 23 |** The output curve of the same device shown in Figure 4a. The linear region at small drain bias indicates the contact resistance is negligible.



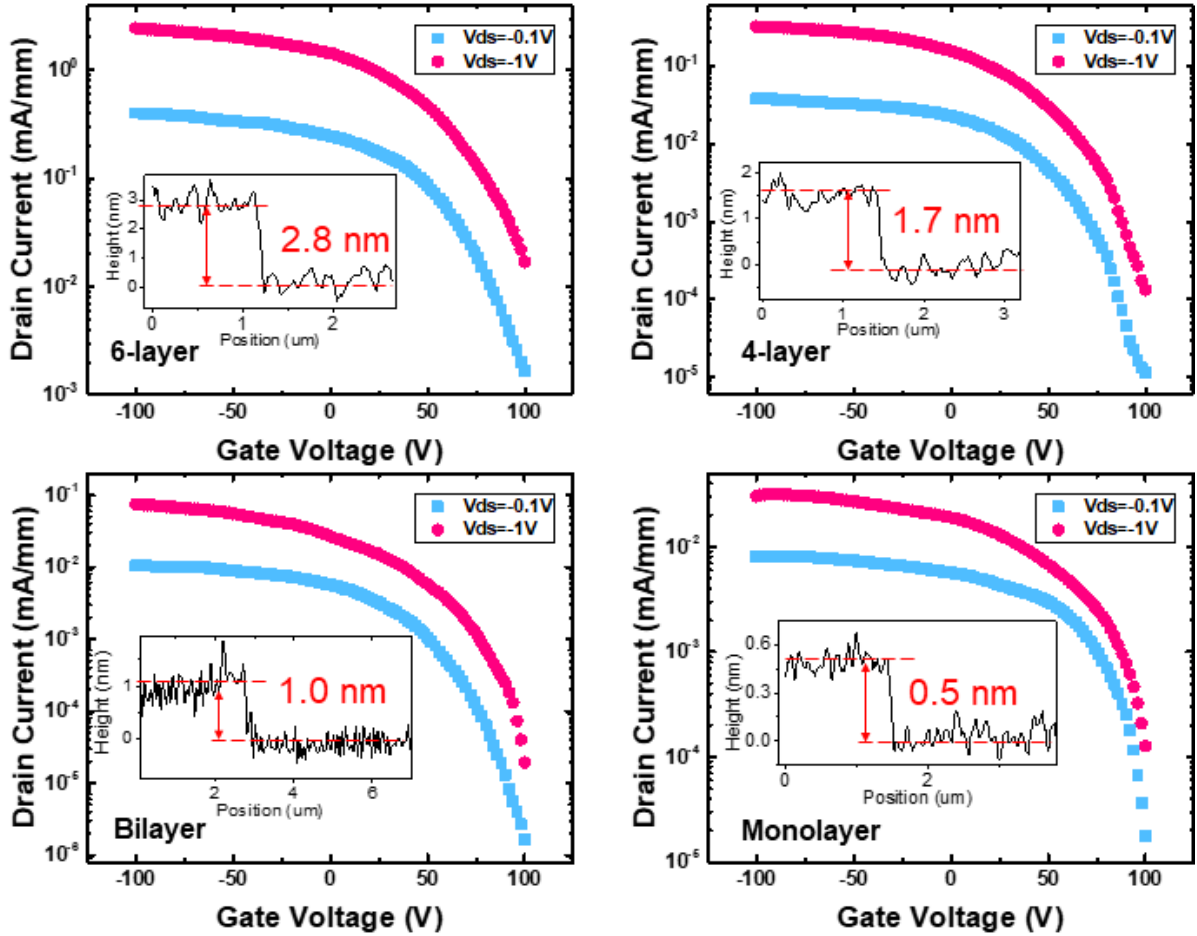

**Supplementary Figure 24** | Transfer curves of 2D Te transistors fabricated on 2.8 nm (6L), 1.7 nm (4L), 1.0 nm (2L) and 0.5 nm (1L) flakes, respectively.



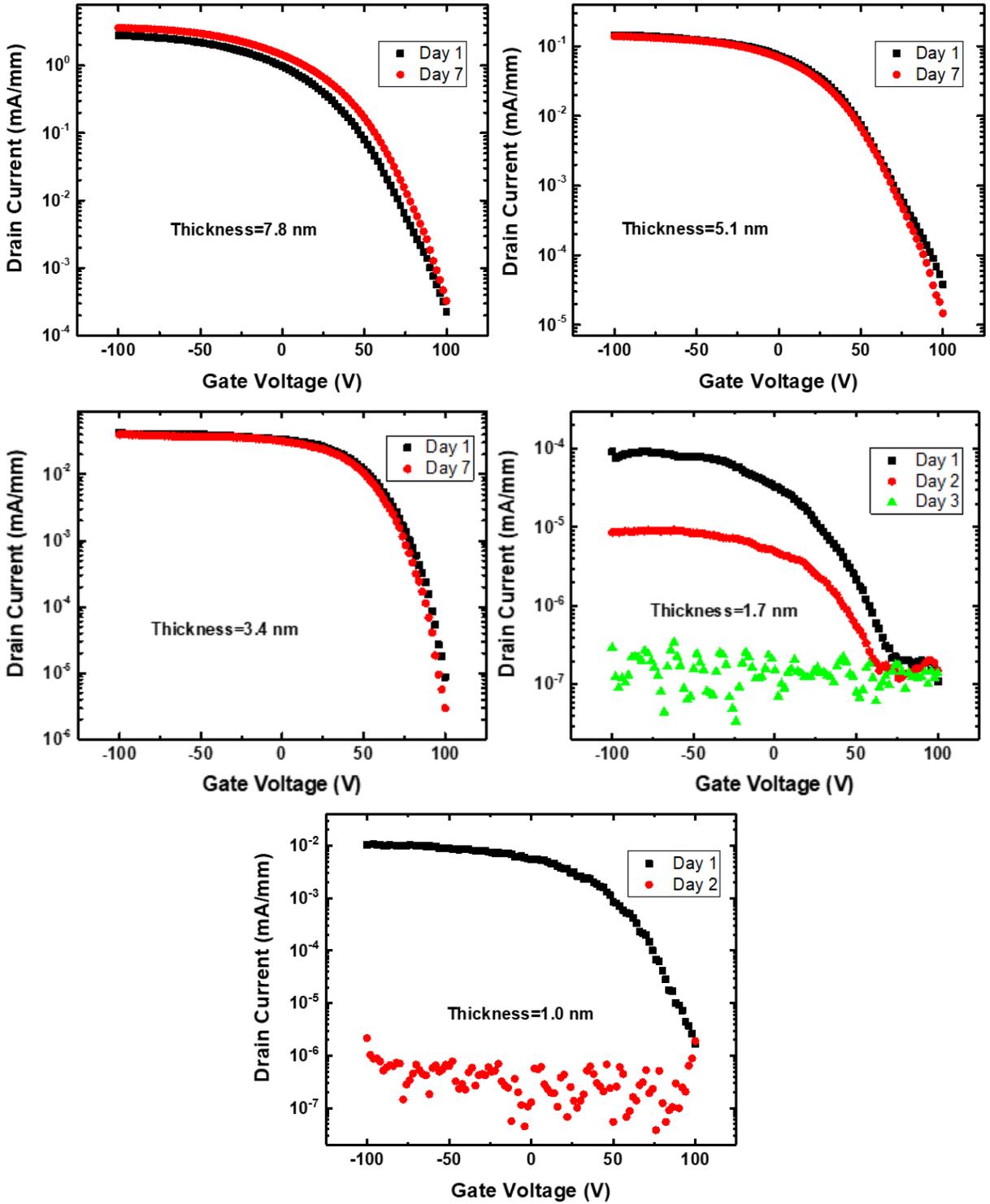

**Supplementary Figure 25** | Stability characterization for 2D Te devices with thickness of 7.8 nm, 5.1 nm, 3.4 nm, 1.7 nm, and 1.0 nm measured as-fabricated and after a controlled period without encapsulation in the air.



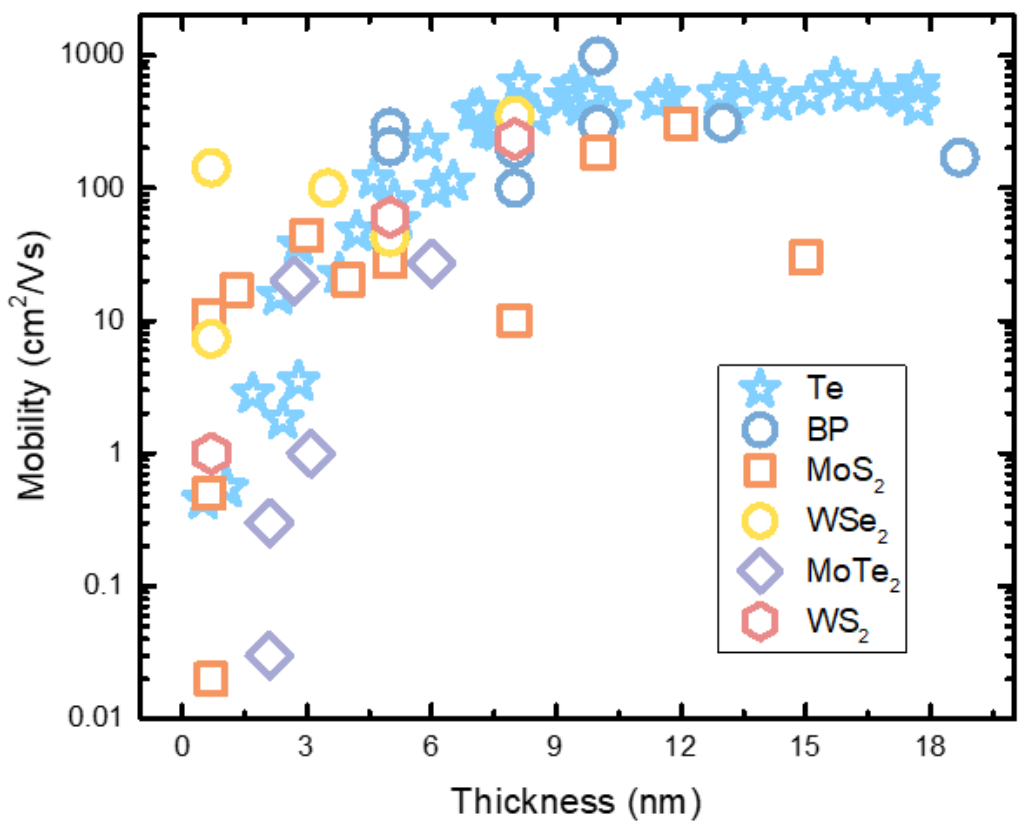

**Supplementary Figure 26** | A comparison of mobility values between 2D Te and other 2D semiconductors in different thickness ranges. The data for other 2D materials are extracted from the references listed in the Supplementary Table 1.



**Supplementary Table 1:** Summary and Comparison of mobility values between 2D Te and other 2D semiconductors in different thickness ranges.[1]

|  | Reference | Thickness (nm) | Mobility (cm$^2$/Vs) | Comments |
|---|---|---|---|---|
| **BP** | ACS Nano, 2014, 8 (4),4033 | 5 | 286 |  |
|  | Nat. Nanotechnology, 2014, 9, 372 | 10 | 984 |  |
|  |  | 8 | 197 |  |
|  |  | 5 | 55 |  |
|  | Nat. Commun., 2014, 5, 4458 | 5 | 205 |  |
|  | Nano Lett., 2014, 14 (6),3347 | 8 | 100 |  |
|  | Appl. Phys. Lett., 2014, 104, 103106 | 10 | 300 |  |
|  | ACS Nano, 2014, 8 (10), 10035 | 18.7 | 186 | Pd contacts |
|  | Nano Lett., 2015, 15(8), 4914 | 0.55 | ~1 | low-T (10K), h-BN capping |
|  | Nano Lett., 2015, 15 (3), 1883 | 13 | 310 |  |
| **MoS$_2$** | IEEE Electron Device Lett., 2013, 34(10),1328 | 4 | 20.4 | Molecular Doping |
|  | Proc. Natl. Acad. Sci., 2005, 102(30), 10451 | 0.65 | 0.5~3 |  |
|  | J. Appl. Phys., 2007, 101, 014507 | 8~40 | 10~50 |  |
|  |  | 35 | 40 |  |
|  | Appl. Phys. Lett. 102, 142106 (2013) | 1.3 | 17 | CVD |
|  | Adv. Mater., 2012, 24(17), 2320 | 0.7 | 0.02 | CVD |
|  | Nano Lett., 2013, 13 (1), 100 | 10 | 184 | Scandium Contacts |
|  | ACS Nano, 2013, 7 (6), 5446 | 15 | 30 | high-k dielectric |
|  | Appl. Phys. Lett., 2014, 104, 093106 | 0.65 | 11 | Molybdenum contacts |
|  | Appl. Phys. Lett., 2013, 102, 123105 (2013) | 12 | 306 | Four-terminal devices |
|  | ACS Nano, 2012, 6 (10), 8563 | 5 | 28 |  |

---

[1] The prototypical 2D Te transistor in this work is merely a demonstration of its exciting electrical properties, whereas most of the state-of-the-art 2D transistors have adopted matured techniques such as doping, dielectric/contact engineering. The summary in **Table 1** shows that 2D Te device has a good all-around figure of merits compared to existing 2D semiconductors. The field-effect mobility of 2D Te has outnumbered almost all *p*-type 2D materials and most of the conventional ultra-thin-body (UTB) semiconductors (Si, GaAs, InAs, etc.). The I$_{on}$ current also surpasses most of its competitors. This is significant in the sense that high-performance PMOS is in principle more challenging to implement than NMOS because holes have lower mobility and larger effective mass than electrons.



| | | | | |
|---|---|---|---|---|
| | ACS Nano, 2013, 7 (7), 5870 | 3 | 44 | |
| **WSe$_2$** | Nano Lett. 2012, 12, (7), 3788 | 0.7 | 250 | NO$_2$ doping, high-k dielectric |
| | Nano Lett., 2013, 13 (5), 1983 | 0.7 | 142 | electrons, Indium contacts |
| | ACS Nano, 2014, 8 (7), 7180 | 0.7 | 30 | polymer electrolyte gating |
| | Sci. Rep., 2015, 5, 8979 | 8 | 350 | |
| | Nano Lett., 2016, 16 (3), 1896 | 3.5 | 200 | Nb-doped contacts |
| | ACS Nano, 2014, 8 (8), 8653 | 0.7 | 7.3 | CVD |
| | Nano Lett., 2015, 15 (8), 4928 | 5 | 27.4 | electrons |
| | | 5 | 42.6 | holes |
| **MoTe$_2$** | ACS Nano, 2014, 8 (6), 5911 | 2.7 | 20 | holes |
| | | 6 | 27 | electrons |
| | J. Am. Chem. Soc., 2015, 137 (37), 11892 | 3.1 | 1 | |
| | Adv. Mater., 2014, 26(20), 3263 | 2.1 | 0.03 | holes |
| | | 2.1 | 0.3 | electrons |
| **WS$_2$** | ACS Nano, 2014, 8 (8), 8174 | 0.7 | 1 | |
| | Nano Lett., 2014, 14 (11), 6275 | 5 | 60 | Chloride Molecular Doping |
| | ACS Nano, 2014, 8 (10), 10396 | 6~8 | 234 | |
| | Sci. Rep., 2014, 4, 5219 | 42 | 16 | |